\documentclass[twocolumn,showpacs,amsfonts,aps,prc,nofootinbib,floatfix]{revtex4-1}

\usepackage{mathrsfs}
\usepackage{epsfig}
\usepackage{amssymb}
\usepackage{amsfonts}
\usepackage{latexsym}
\usepackage{amsthm}

\usepackage{bm}
\usepackage{graphicx}
\usepackage{amsmath}
\usepackage{hyperref}

\usepackage{ulem}
\usepackage{color}

\begin{document}


\title{Collision Energy Dependence of Viscous Hydrodynamic Flow in Relativistic Heavy-Ion Collisions}

\author{Chun Shen}
\author{Ulrich Heinz}
\affiliation{Department of Physics, The Ohio State University,
  Columbus, Ohio 43210-1117, USA}
  
\begin{abstract}
Using a (2+1)-d viscous hydrodynamical model, we study the dependence of flow observables on the collision energy ranging from $\sqrt{s}=7.7$\,$A$\,GeV at the Relativistic Heavy Ion Collider (RHIC) to $\sqrt{s}=2760$\,$A$\,GeV at the Large Hadron Collider (LHC). With a realistic equation of state, Glauber model initial conditions and a small specific shear viscosity $\eta/s = 0.08$, the differential charged hadron elliptic flow $v_2^\mathrm{ch}(p_T,\sqrt{s})$ is found to exhibit a very broad maximum as a function of $\sqrt{s}$ around top RHIC energy, rendering it 
almost independent of collision energy for $39 \le\sqrt{s} \le 2760$\,$A$\,GeV. Compared to ideal fluid dynamical simulations, this ``saturation" of elliptic flow is shifted to higher collision energies by shear viscous effects. For color-glass motivated MC-KLN initial conditions, which require a larger shear viscosity $\eta/s = 0.2$ to reproduce the measured elliptic flow, a similar ``saturation" is not observed up to LHC energies, except for very low $p_T$. We emphasize that this ``saturation" of the elliptic flow is not associated with the QCD phase transition, but arises from the interplay between radial and elliptic flow which shifts with $\sqrt{s}$ depending on the fluid's viscosity and leads to a subtle cancellation between increasing contributions from light and decreasing contributions from heavy particles to $v_2$ in the $\sqrt{s}$ range where $v_2^\mathrm{ch}(p_T,\sqrt{s})$ at fixed $p_T$ is maximal. By generalizing the definition of spatial eccentricity $\epsilon_x$ to isothermal hyper-surfaces, we calculate $\epsilon_x$ on the kinetic freeze-out surface at different collision energies. Up to top RHIC energy, $\sqrt{s}=200$\,$A$\,GeV, the fireball is still out-of-plane deformed at freeze out, while at LHC energy the final spatial eccentricity is predicted to approach zero.
\end{abstract}


\date{\today}

\maketitle

\section{Introduction}
\label{sec1}

The first decade of experiments at the Relativistic Heavy Ion Collider (RHIC) has established the existence of a strongly coupled Quark-Gluon Plasma (sQGP), a new state of nuclear matter with partonic degrees of freedom. One of the major goals of heavy-ion collision experiments right now is to explore the QCD phase diagram. The recent Beam Energy Scan (BES) program \cite{Caines:2009yu,Kumar:2011de,Pandit:2011hf,Shi:2011ad,Kumar:2011us,Schmah:2011zz} at RHIC is motivated by searching for the phase boundary between normal nuclear matter and sQGP as well as for the theoretically predicted QCD critical point \cite{Alford:2007xm,Fodor:2001pe,Ejiri:2003dc,Gavai:2004sd}. The BES program at RHIC together with Pb+Pb collisions at the Large Hadron Collider (LHC) provide us with a unique opportunity to study systematically  the collision energy dependence of relativistic heavy-ion collision observables.

The study of collective flow in relativistic heavy ion collisions has the potential to offer insights into properties of the produced matter. Anisotropic flows, especially the elliptic flow $v_2$, are widely studied in heavy-ion collisions. In the mid-rapidity region, the dependence of elliptic flow on transverse momentum and collision energy are crucial for our understanding of the properties of sQGP. Elliptic flow can provide information about the specific shear viscosity, which controls the conversion efficiency of anisotropic spatial pressure gradients to momentum anisotropies in a hydrodynamic description, and about the equation of state of the matter created at early times \cite{Teaney:2003kp,Lacey:2006pn,Romatschke:2007mq,Song:2007fn,Song:2008si,Luzum:2008cw,Luzum:2009sb,Song:2010mg,Aamodt:2010pa,Luzum:2010ag,Lacey:2010ej,Bozek:2010wt,Hirano:2010jg,Schenke:2010rr,Schenke:2011tv,Song:2011qa,Shen:2010uy,Shen:2011eg}. Recent measurements of higher order anisotropic flow coefficients, $v_n (n\ge3)$
have generated strong interest due to their ability to provide additional constraints on initial conditions \cite{Alver:2010gr,Adare:2011tg,Alver:2010dn,Sorensen:2011fb,ALICE:2011vk,CMSflow,Steinberg:2011dj,Petersen:2010cw,Qin:2010pf,Luzum:2010sp,Xu:2011fe,Luzum:2011mm,Qiu:2011hf,Chaudhuri:2011qm,Schenke:2011bn,Lacey:2010hw,Shen:2011zc,Qiu:2011fi}. 

In \cite{Kestin:2008bh} the collision energy dependence of particle transverse momentum spectra and elliptic flow coefficients were studied using (2+1)-d ideal hydrodynamics with longitudinal boost invariance and a bag-model equation of state. In this work, we revisit this problem using more realistic (2+1)-d viscous hydrodynamics coupled with a modern lattice QCD based equation of state \cite{Shen:2010uy,Huovinen:2009yb}. We also study the differences between the two most popular initial conditions obtained from Monte Carlo versions of the Glauber (MC-Glauber) and Kharzeev-Levin-Nardi (MC-KLN) models \cite{Kharzeev:2001yq,Hirano:2005xf,Drescher:2006pi,Hirano:2009ah,Heinz:2009cv}.

Our work has some limitations which must be kept in mind before comparing them with experimental data. As the collision energy decreases, the Bjorken assumption of longitudinal boost invariance will gradually break down \cite{Morita:2002av}. Furthermore, since the fireball will spend less time in the QGP phase, the hadronic phase becomes more important and occupies a larger part in its dynamical history. Curing these two major shortcomings will require (3+1)-d viscous hydrodynamic simulations \cite{Schenke:2010rr, Vredevoogd:2012ui} coupled with a microscopic hadronic afterburner \cite{Bass:1998ca,Song:2010mg,Song:2010aq}. The present work does not aim at extracting precise information of QGP transport properties from a comparison with experimental data. Its main purpose is to expose systematic quantitative trends in observables as a function of collision energy in the relativistic heavy-ion collisions. 

In the next section, we will describe the setup of our models and discuss our parametrization of the initial conditions as a function of $\sqrt{s}$. In Sec III, we will present trends for the transverse momentum spectra and differential elliptic flow for charged hadrons as $\sqrt{s}$ increases from 7.7\,$A$\,GeV to 2760\,$A$\,GeV. Identified particle spectra and their elliptic flow $v_2$ will be discussed in Sec. IV. In Sec. V, we generalize the definition of the spatial eccentricity to an isothermal hyper-surface. Based on this generalized formulation, we perform a shape analysis on the final kinetic freeze-out surface and study the dependence of the final eccentricity on $\sqrt{s}$. Sec. VI is devoted to some concluding remarks.

\section{Evolution of charged hadron multiplicity and total elliptic flow}
\label{sec2}

In this work, we employ the (2+1)-dimensional viscous hydrodynamic model {\tt VISH2+1} which implements boost-invariance in the longitudinal direction \cite{Song:2007fn}.  Similar to past work \cite{Luzum:2008cw,Hirano:2009ah,Luzum:2009sb,Heinz:2009cv,Qiu:2011iv,Song:2010mg,Qiu:2011hf}, we use two different types of initializations taken from the MC-Glauber and MC-KLN models to generate initial entropy density profiles.  Over one million Monte Carlo events are generated and sorted into collision centrality bins according to their number of participant nucleons. Each event is re-centered to the beam axis and rotated in the transverse plane such that its minor axis aligns with the impact parameter. Then we average the events to obtain a smooth average initial entropy density for each centrality bin.

In the MC-Glauber runs we take for the specific shear viscosity the value $\eta/s = 0.08$ since this value was shown in \cite{Schenke:2011tv, Schenke:2011bn,ALICE:2011vk,Qiu:2011hf} to provide a reasonable description of the charged hadron $v_2(p_T)$ and $v_3(p_T)$ data measured by the RHIC and the LHC experiments \cite{Adare:2011tg,Sorensen:2011fb,ALICE:2011vk,CMSflow,Steinberg:2011dj}. For the initial entropy density we make the two-component ansatz
\begin{equation}
  s(\bm{r}_\perp;b)=\kappa \left( \frac{1-x}{2} n_{_\mathrm{WN}} (\bm{r}_\perp;b) 
                                                + x  n_{_\mathrm{BC}}(\bm{r}_\perp;b)\right),
\label{eq1}
\end{equation}
with a wounded nucleon (WN) to binary collision (BC) mixing ratio $x$ that is adjusted to reproduce the measured centrality dependence of the final charged hadron multiplicity density $dN_\mathrm{ch}/d\eta$. For Pb+Pb collisions at $\sqrt{s}=2760$\,$A$\,GeV we use $x=0.118$ as determined in \cite{Qiu:2011hf}. For Au+Au collisions at RHIC energies, we determine $\kappa$ and $x$ in Eq. (\ref{eq1}) by a two-parameter fit to the RHIC data at $\sqrt{s}=200$\,$A$\,GeV \cite{Abelev:2008ez,Back:2004dy,Alver:2008ck,Back:2002uc,Adler:2004zn} obtaining $x=0.14$. For RHIC collisions at lower $\sqrt{s}$ we keep the mixing ratio fixed\footnote{The main reason of keeping the mixing ratio fixed is because the measured centrality dependence of charged multiplicity for the lower energy runs has not yet been published. We do not expect qualitative changes to the conclusions drawn in this paper once our assumed values will be replaced by actual measurements.} at $x=0.14$,  tuning only the normalization factor $\kappa$ to reproduce the charge multiplicity in the 0-5\% most central collisions. For $\sqrt{s}=63$\,$A$\,GeV, the desired charged multiplicity is taken from experiment \cite{Abelev:2008ez}. For $\sqrt{s} < 63$\,$A$\,GeV, we presently lack experimental information and therefore use the empirical formula \cite{Kestin:2008bh}
\begin{equation}
\frac{dN_\mathrm{ch}}{d \eta} = 312.5\, \mathrm{log}_{10} \sqrt{s} - 64.8.
\label{eq2}
\end{equation}
The actually employed final charged multiplicities are listed in Table \ref{table1}.

\begin{table}
\begin{tabular}{|c|c|c|c|}
\hline
$\sqrt{s}$ (A GeV) & $T_0$ (MeV) & $\tau_\mathrm{f}{-}\tau_0$ (fm/$c$) & $dN_\mathrm{ch}/d\eta$ \\ \hline
AuAu@ 7.7    & 269.2/233.7  & 9.3/9.1	& 212.3/212.1  \\ \hline
AuAu@ 11.5  & 287.5/252.0  & 10.0/9.8	& 266.7/266.4  \\ \hline
AuAu@ 17.7  & 304.8/269.8  & 10.5/10.3 	& 325.3/324.9  \\ \hline
AuAu@ 19.6  & 308.7/274.3  & 10.6/10.4 	& 339.2/338.8  \\ \hline
AuAu@ 27     & 320.1/286.4  & 10.9/10.7 	& 382.9/382.1  \\ \hline
AuAu@ 39     & 332.2/298.9  & 11.2/11.0  	& 432.7/432.3  \\ \hline
AuAu@ 63     & 341.1/306.4  & 11.4/11.2  	& 472.0/472.9  \\ \hline
AuAu@ 200   & 378.6/347.0  & 12.2/12.1 	& 661.9/690.0  \\ \hline
PbPb@ 2760 & 485.2/443.9  & 14.2/14.2 	& 1575.7/1597.2 \\ \hline
\end{tabular}
\caption{The initial temperature at the center of the fireball, fireball lifetime and final charged hadron multiplicity of $0$-$5\%$ most central collisions are listed.  The results on the left are from MC-Glauber initial conditions with $\eta/s=0.08$, the right are for MC-KLN with $\eta/s=0.2$. }
\label{table1}
\end{table}

The MC-KLN calculations are done using a Monte-Carlo sample of initial profiles with identical properties as those used in \cite{Shen:2011eg}. These initial MC-KLN profiles were evolved hydrodynamically with a larger viscosity $\eta/s=0.2$ to compensate for the larger initial eccentricities. Again, this choice was shown to yield a good overall description of the measured transverse momentum spectra and elliptic flow in 200\,$A$\,GeV Au-Au collisions at RHIC \cite{Shen:2011eg} and gave an impressively accurate prediction \cite{Shen:2011eg,Heinz:2011kt} for the unidentified and identified charged hadron spectra and elliptic flows in 2760\,$A$\,GeV Pb-Pb collisions at the LHC \cite{Aamodt:2010pa,ALICE:2011vk}. The large $\eta/s$ of 0.2 fails, however, to reproduce the large $v_3$ measured in Pb+Pb collisions at the LHC \cite{Qiu:2011hf, ALICE:2011vk}. In the MC-KLN model, the initially produced gluon density profile is controlled by the dependence of the saturation scale $Q_s$ on the position $\bm{x_\perp}$ in the transverse plane. For a nucleus with mass number $A$ it is given by \cite{Hirano:2009ah}
\begin{equation}
Q^2_{s,A}(x; \bm{x_\perp}) = Q^2_{s,0} \frac{T_A(\bm{x}_\perp)}{T_{A,0}}\left(\frac{x_0}{x}\right)^\lambda,
\end{equation}
where $T_A (\bm{x}_\perp)$ is the nuclear thickness function. We use the same parameter set($Q^2_{s,0} = 2$\,GeV$^2$, $T_{A,0} = 1.53$\,fm$^{2}$, $\lambda = 0.28$ and $x_0 = 0.01$) as proposed in Ref. \cite{Hirano:2009ah}. For Au+Au at 200\,$A$\,GeV and Pb+Pb at 2760\,$A$\,GeV the normalization constant for the initial entropy density was determined by an overall fit to the centrality dependence of $\frac{dN_\mathrm{ch}}{d\eta}$. These best fits result in slightly different $\frac{dN_\mathrm{ch}}{d\eta}$ values for the 0-5\% most central collisions than obtained for the corresponding MC-Glauber cases (see Table~\ref{table1}). At lower energies, the normalization factor was again fixed to reproduce the desired charged hadron multiplicity density $dN_\mathrm{ch}/d\eta$ for the 0-5\% most central collisions for all $\sqrt{s}$ (see Table \ref{table1}). 

Starting from an assumed thermalization time $\tau_0 = 0.6$ fm/$c$ \cite{Shen:2010uy,Shen:2011eg,Qiu:2011hf}, both the MC-Glauber and MC-KLN initial conditions are evolved hydrodynamically using the lattice QCD based equation of state (EOS) s95p-PCE \cite{Shen:2010uy, Huovinen:2009yb}. This EOS accounts for chemical freeze-out at $T_\mathrm{chem} = 165$\,MeV before thermal decoupling which is taken to occur along an isothermal  surface of temperature $T_\mathrm{dec}=120$\,MeV. We convert the hydrodynamic output along the kinetic decoupling surface into final hadron distributions using the Cooper-Frye prescription \cite{Cooper:1974mv}. Strong resonance decays are taken into account up to 2 GeV in particle mass. 

We point out that we keep the value of the specific shear viscosity $\eta/s$ unchanged as we go to lower collision energies. As the highly viscous hadronic phase becomes more and more important at lower collision energies, viscous hydrodynamic simulations with temperature independent $\eta/s$ will eventually break down. Worse, larger $\eta/s$ values in the hadronic phase jeopardize the validity of the viscous hydrodynamic approach altogether \cite{Shen:2011kn}. In this study, we are not trying to extract the temperature dependence of $\eta/s$ from a serious comparison with experimental data; our goal is to present a systematic study of the $\sqrt{s}$ dependence of hydrodynamic variables.  For this reason, we run viscous hydro all the way down to $\sqrt{s}= 7.7$\,$A$\,GeV with constant $\eta/s$ compared to \cite{Kestin:2008bh}, our simulations are more realistic by including viscous effects in the hydrodynamic evolution and using a better EOS.\footnote{It should be noted that our EOS assumes zero net baryon density -- an assumption that is untenable in the lower half of the collision energy range explored here. To include effects from non-zero baryon density would require an upgrade of {\tt VISH2{+}1} to solve additionally for the space-time evolution of the conserved baryon current. This is important for the correct prediction of the final baryon and meson abundances at lower $\sqrt{s}$ which our present code can not achieve. However, what matters for the evolution of radial and elliptic flow is the stiffness of the EOS, embodied by the pressure (whose gradients supply the hydrodynamic acceleration) and its relation to the energy density (inertia) of the fluid, $p(e,n)$. Since, for not too large baryon densities $n$, this relation depends on $n$ only very weakly \cite{Kolb:2000sd}, the use of a baryon-free EOS is expected to work well for the systematic flow study presented here.}
Also, we here study two different initialization models and include (at least on average) the effects of event-by-event fluctuations whereas in \cite{Kestin:2008bh} an optical Glauber model was used for initialization which gives too small eccentricities in most central collisions. We will see that the different $\sqrt{s}$ dependence from the two initialization models will help us to further distinguish between the two initialization models.

In Table \ref{table1} we have summarized the global variables for our hydrodynamic simulations. At higher collision energies the evolution starts with a higher peak initial temperature, thus probing the nuclear matter at higher temperature and resulting in a longer lifetime of the fireball. At LHC energy we find a peak temperature that is about twice as large as that reached at the lowest collision energies at RHIC, and the lifetime is about 5 fm/$c$ longer. MC-Glauber initial conditions have about 30 MeV higher peak temperatures than MC-KLN ones. This is mostly due to the fact that the specific shear viscosity in the MC-KLN runs is about 2.5 times larger than for MC-Glauber runs, causing stronger viscous heating and larger entropy production during the hydrodynamic evolution. The same final multiplicity $\frac{dN_\mathrm{ch}}{d\eta}$ can thus be reached starting from less initial entropy. A larger specific shear viscosity also helps the system to develop more radial flow  in the transverse plane, by speeding up the equalization between transverse and longitudinal velocity gradients (the latter are initially very large). This larger transverse expansion rate compensates for the viscous heating effects on the lifetime, resulting in a slightly shorter lifetime for the MC-KLN runs.

\begin{figure*}
  \begin{minipage}{0.79\linewidth}
  \begin{tabular}{cc}
  \includegraphics[width=0.49\linewidth,height=5.5cm]{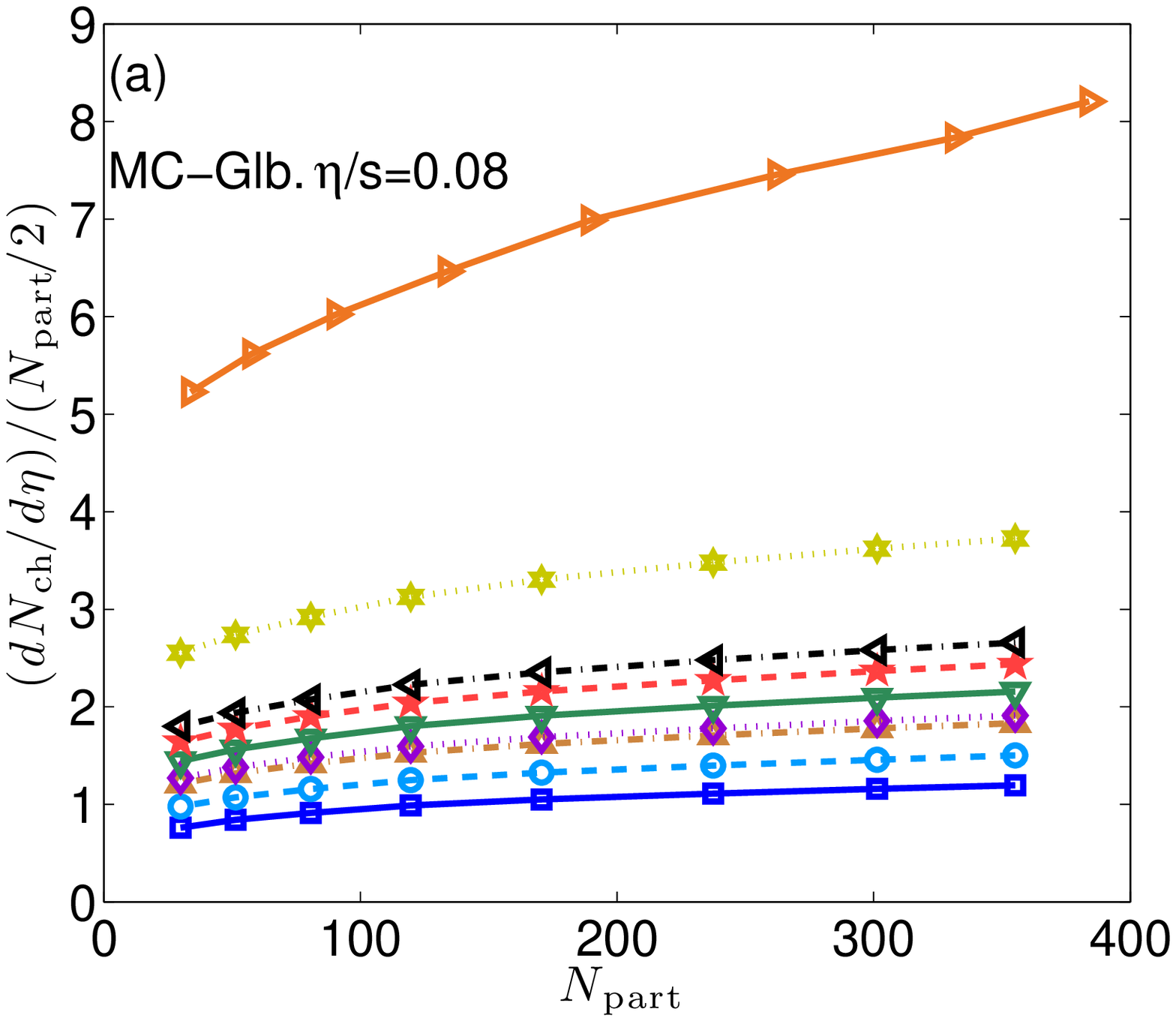} &
  \includegraphics[width=0.49\linewidth,height=5.5cm]{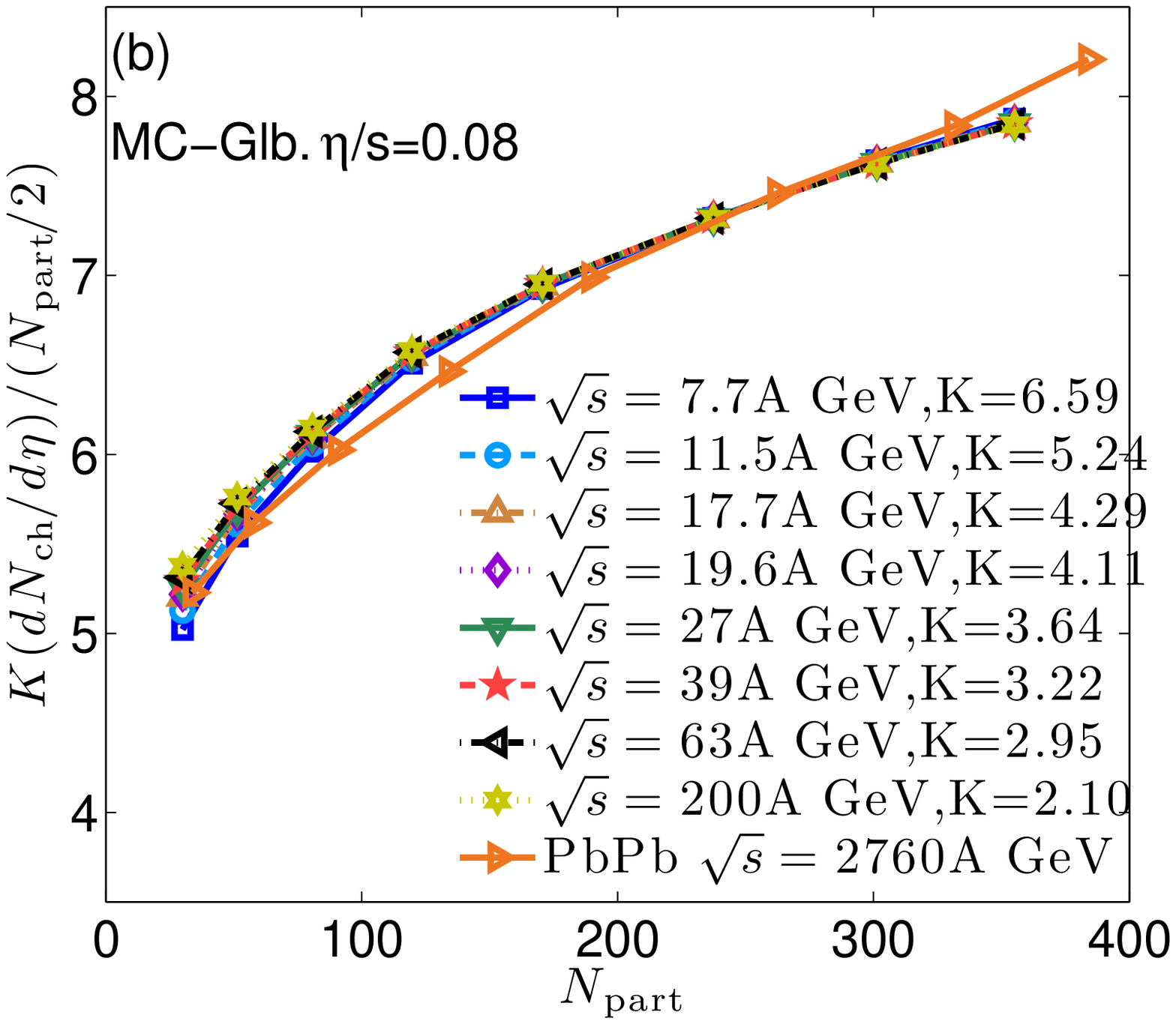} \\
  \includegraphics[width=0.49\linewidth,height=5.5cm]{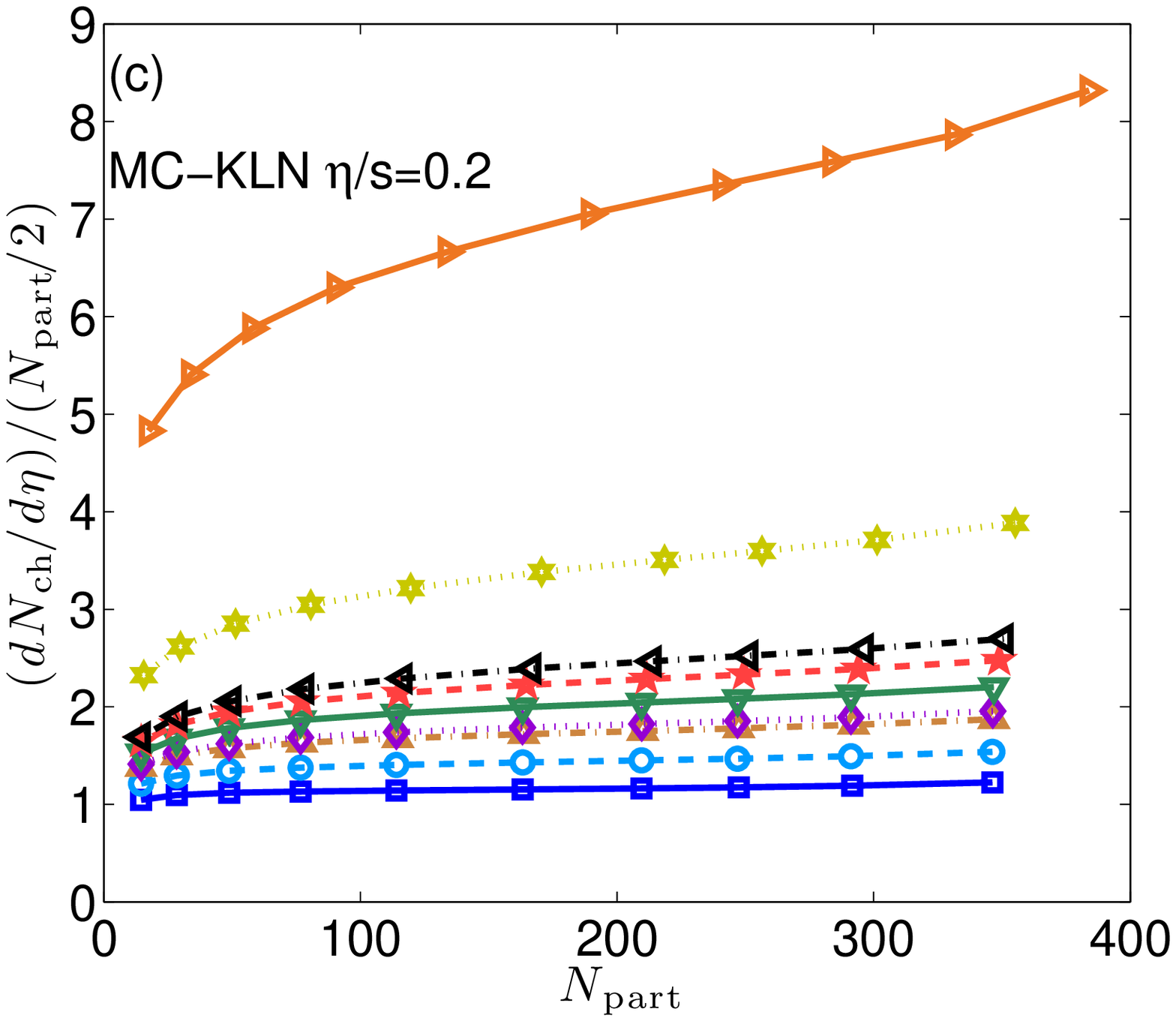} &
  \includegraphics[width=0.49\linewidth,height=5.5cm]{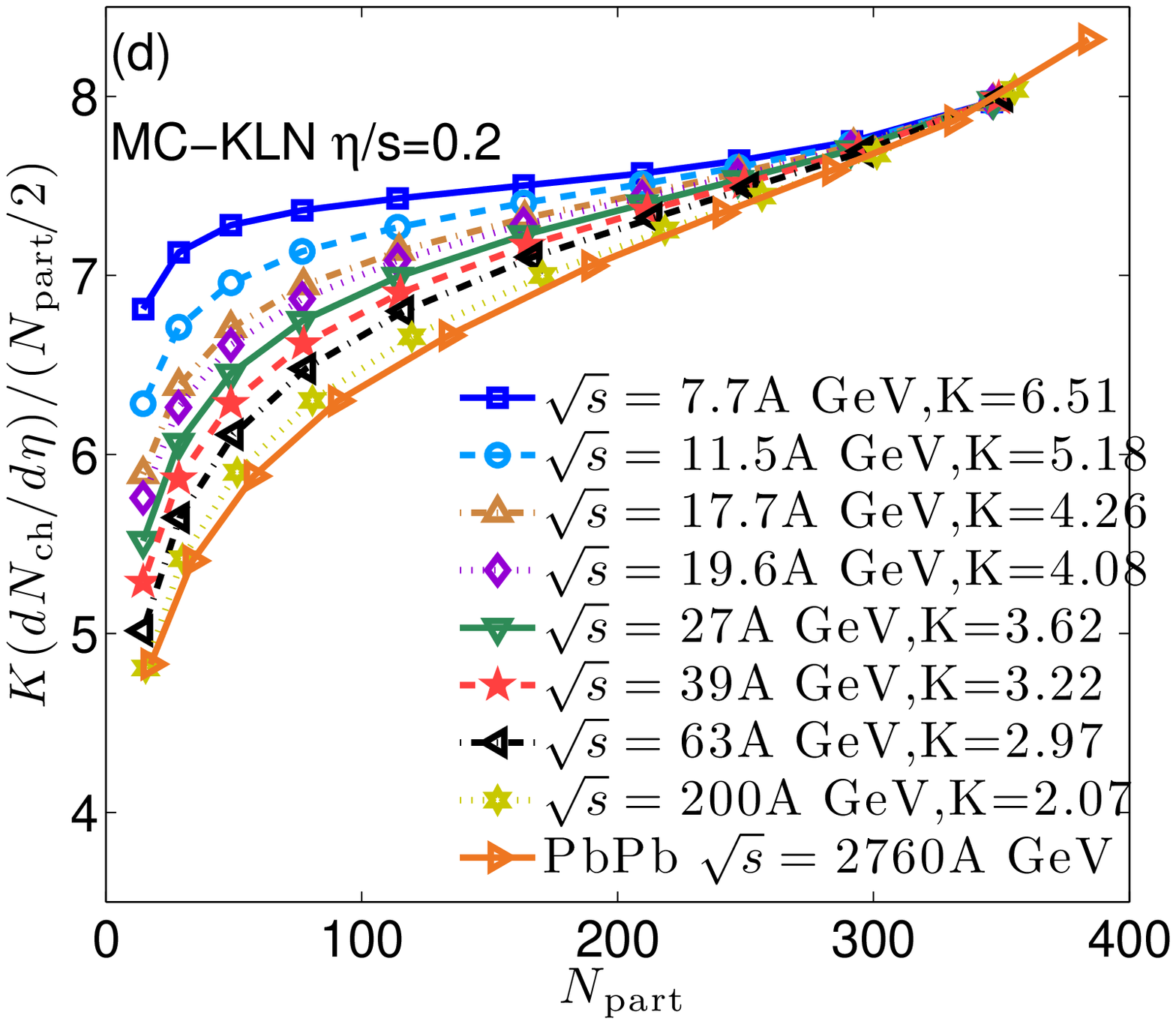}
  \end{tabular}
  \end{minipage}
  \begin{minipage}{0.2\linewidth}
  \caption{{\bf (a):} Centra\-lity dependence of final charged hadron multipli\-city per participant nucleon pair as a function of $N_\mathrm{part}$ for MC-Glauber initial conditions, with collision energies varying from $\sqrt{s}{\,=\,}7.7A$\,GeV to $\sqrt{s}{\,=\,} 2760 A$\,GeV. {\bf (b):} Centrality dependence of $\frac{dN_\mathrm{ch}}{d\eta}$ from the lower energy runs in (a) scaled up to the LHC results, for shape comparison. {\bf (c, d):} Same as (a, b) but for MC-KLN initial conditions.}
  \label{fig1}
 \end{minipage}
\end{figure*}
%
\begin{figure*}
\begin{minipage}{0.79\linewidth}
\begin{tabular}{cc}
  \includegraphics[width=0.49\linewidth,height=5.5cm]{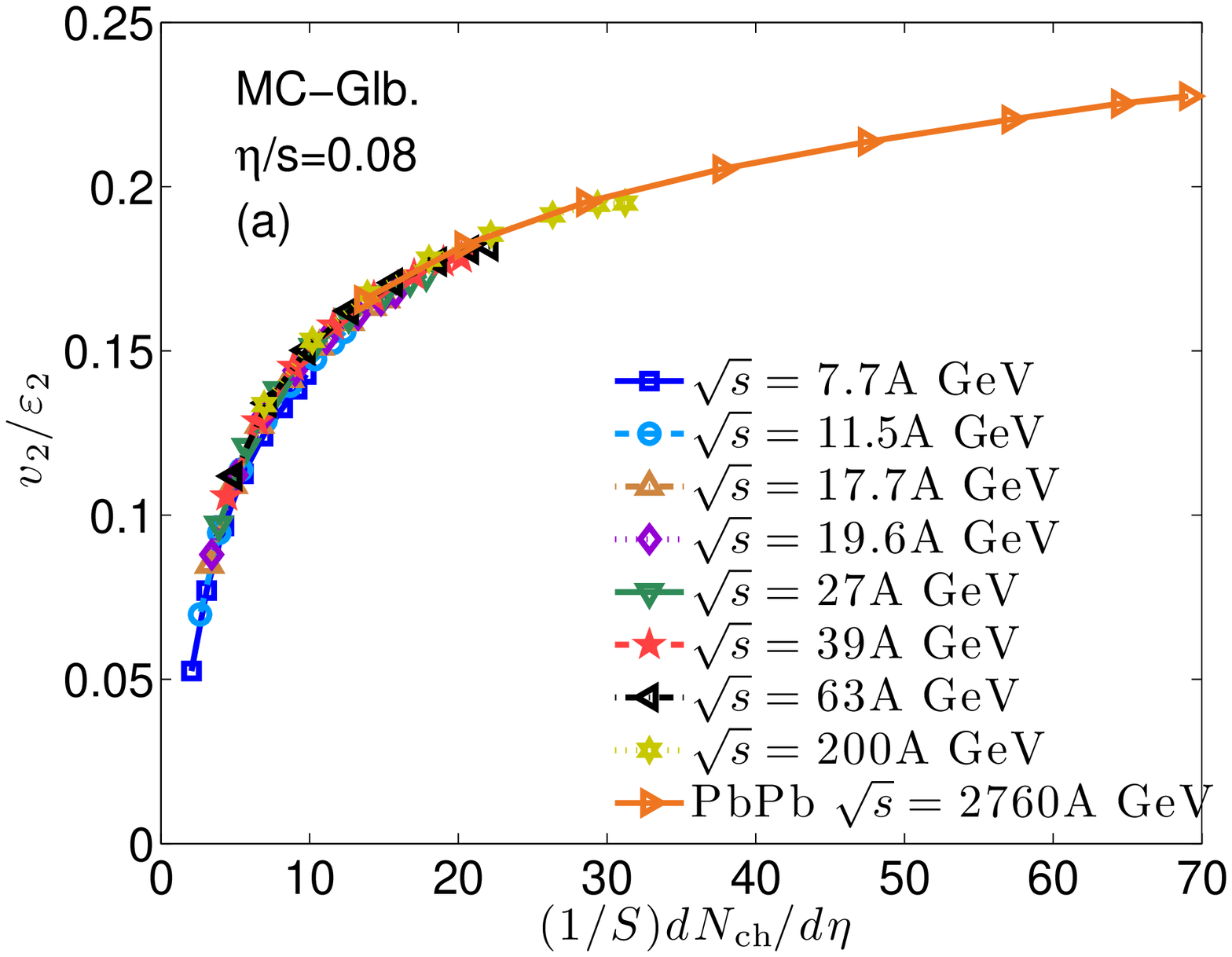} & 
  \includegraphics[width=0.49\linewidth,height=5.5cm]{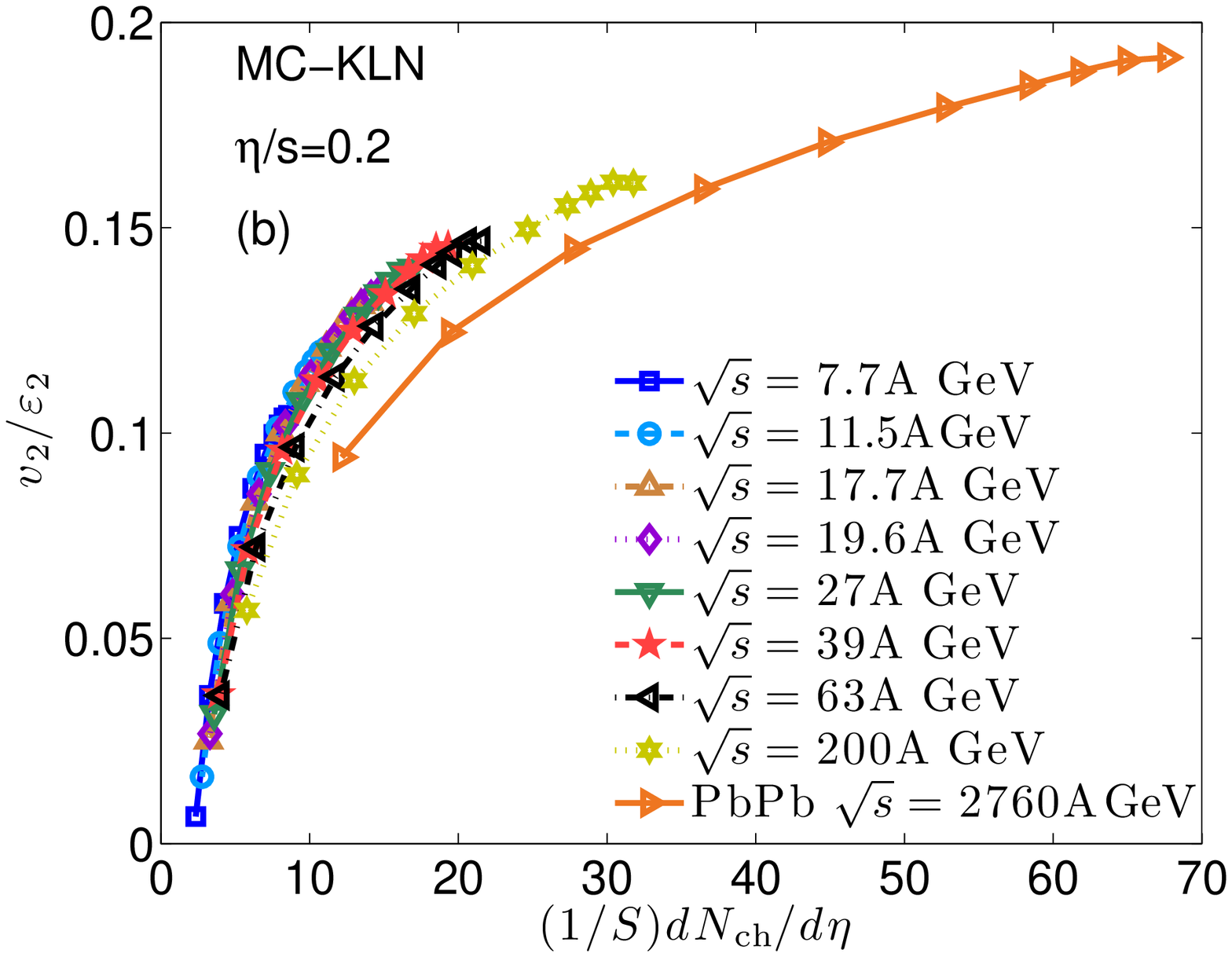}
\end{tabular}
\end{minipage}
\begin{minipage}{0.2\linewidth}
\caption{Eccentricity-scaled $p_T$-integrated $v_2$ plotted as a function of the charged hadron multiplicity density for different collision energies, for MC-Glauber initial conditions with $\eta/s=0.08$ (a) and MC-KLN profiles with $\eta/s=0.2$ (b), respectively. }
\label{fig2}
\end{minipage}
\end{figure*}
%

In Figs. \ref{fig1}(a,c) we show the centrality dependence of the charged hadron multiplicity for both MC-Glauber and MC-KLN models with collision energies from $\sqrt{s}=7.7$ to 2760\,$A$\,GeV.  The reader should note that all results in Fig.~\ref{fig1} account for viscous entropy production during the hydrodynamic evolution. We checked that at LHC and top RHIC energies (top two curves in Figs. 1(a,c)) our results for both initialization models agree well with the experimental data \cite{Abelev:2008ez,Back:2004dy,Alver:2008ck,Back:2002uc,Adler:2004zn,Aamodt:2010pb,Aamodt:2010cz}. Our lower collision energy predictions can in the future be checked against data collected in the RHIC BES program. 

In order to study how the centrality dependence changes with $\sqrt{s}$, we scale in Figs.~\ref{fig1}b,d the lower collision energy results by constant factors to align them with the LHC curve in central (0-10\%) collisions. For the MC-Glauber model we find good $\sqrt{s}$-scaling: the curves almost fall on top of each other. For the low energy runs at RHIC this is, of course, sensitive to the fact that we keep the mixing ratio between the wounded nucleons and binary collisions fixed, and it also reflects the fact that viscous entropy production is small and has little effect on the centrality dependence. On the other hand, for the MC-KLN model the slope of the centrality dependence gets flatter as the collision energy decreases. Only the top RHIC and LHC energy curves approximately fall on top of each other; at lower energy this $\sqrt{s}$-scaling is broken. We found that this tendency originates in the nature of the MC-KLN model itself: Even though viscous entropy production is larger (due to the larger $\eta/s$ used in the MC-KLN runs), its centrality dependence has only a minor effect on the centrality dependence of $\frac{dN_\mathrm{ch}}{d\eta}$ and cannot explain the different shapes of the curves in Figs.~\ref{fig1}b,d. Our MC-KLN calculations thus predict a violation of the $\sqrt{s}$-scaling of the centrality dependence of $\frac{dN_\mathrm{ch}}{d\eta}$ at lower collision energies that is not seen with the MC-Glauber initial conditions. This may help to discriminate experimentally between these models.  

%
\begin{figure*}
  \begin{minipage}{0.79\linewidth}
  \begin{tabular}{cc}
  \includegraphics[width=0.49\linewidth,height=5.5cm]{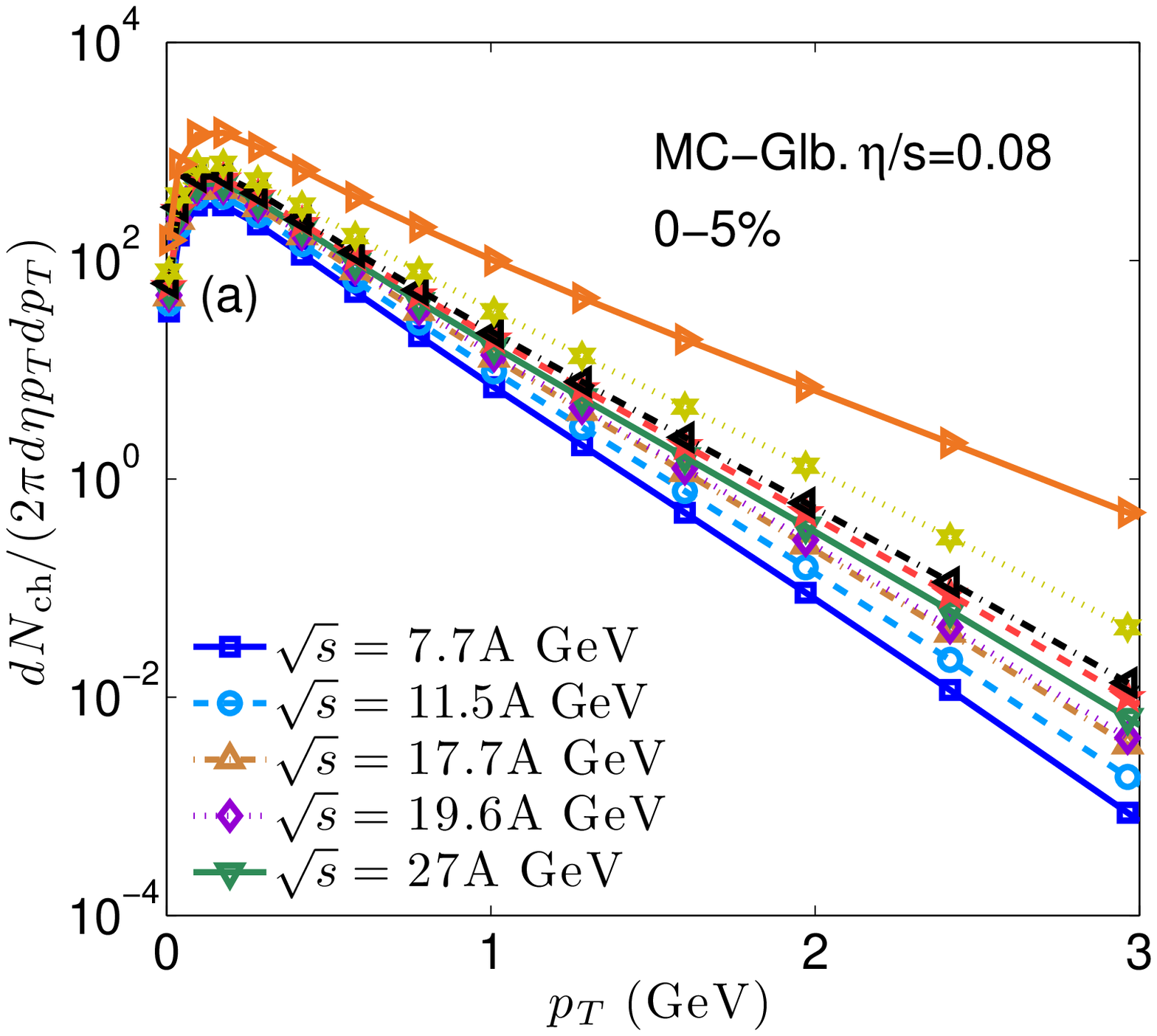} &
  \includegraphics[width=0.49\linewidth,height=5.5cm]{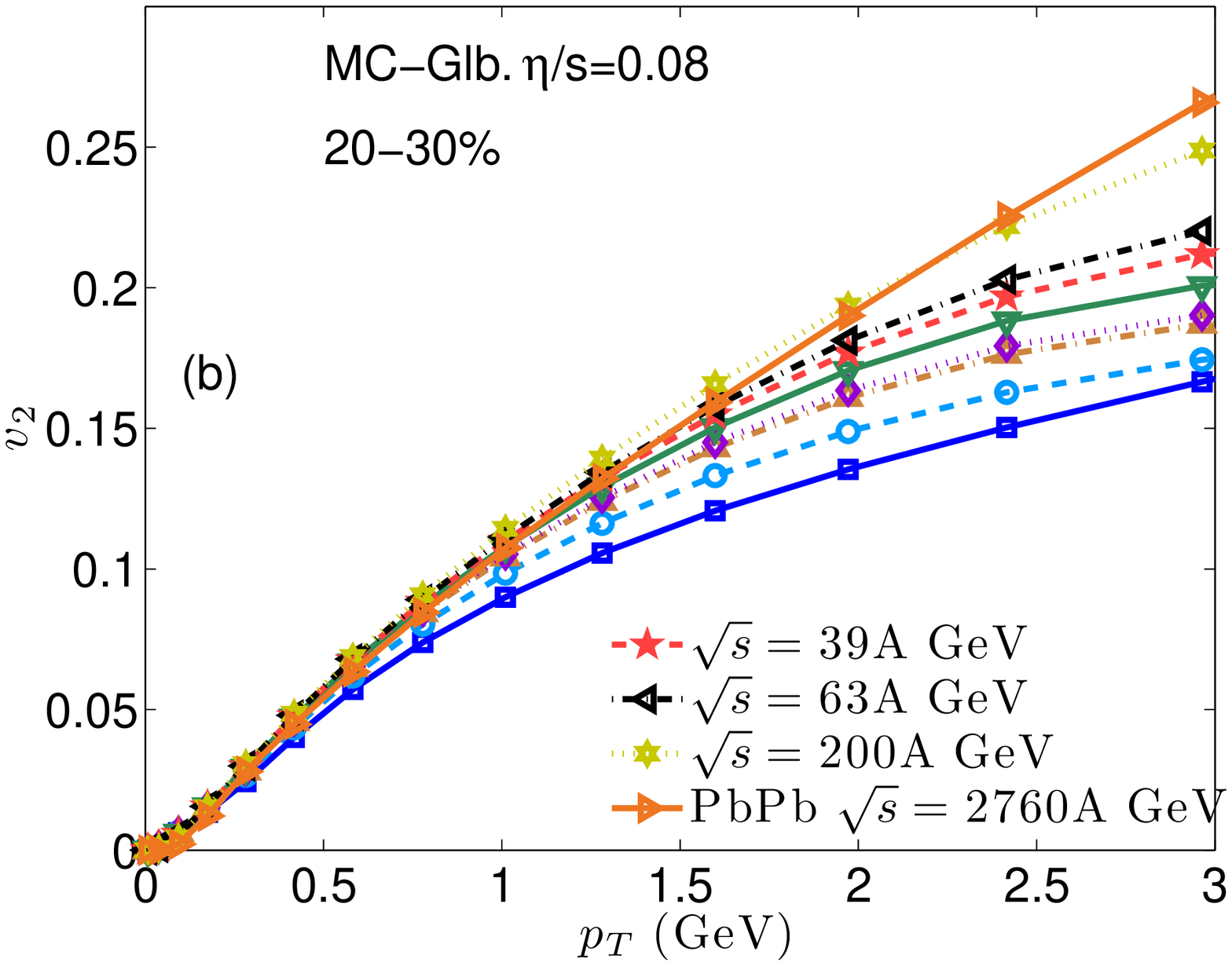} \\
  \includegraphics[width=0.49\linewidth,height=5.5cm]{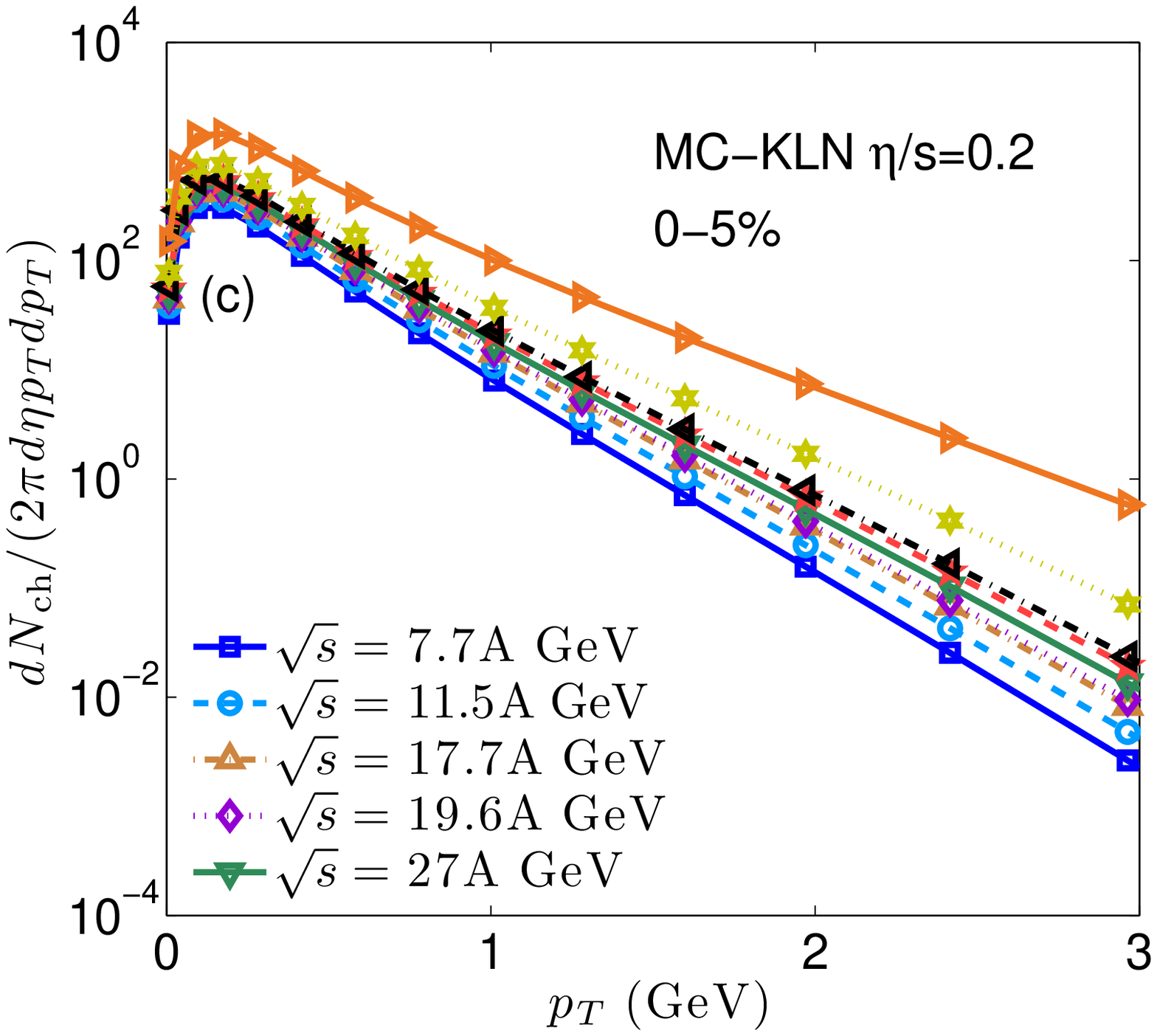} &
  \includegraphics[width=0.49\linewidth,height=5.5cm]{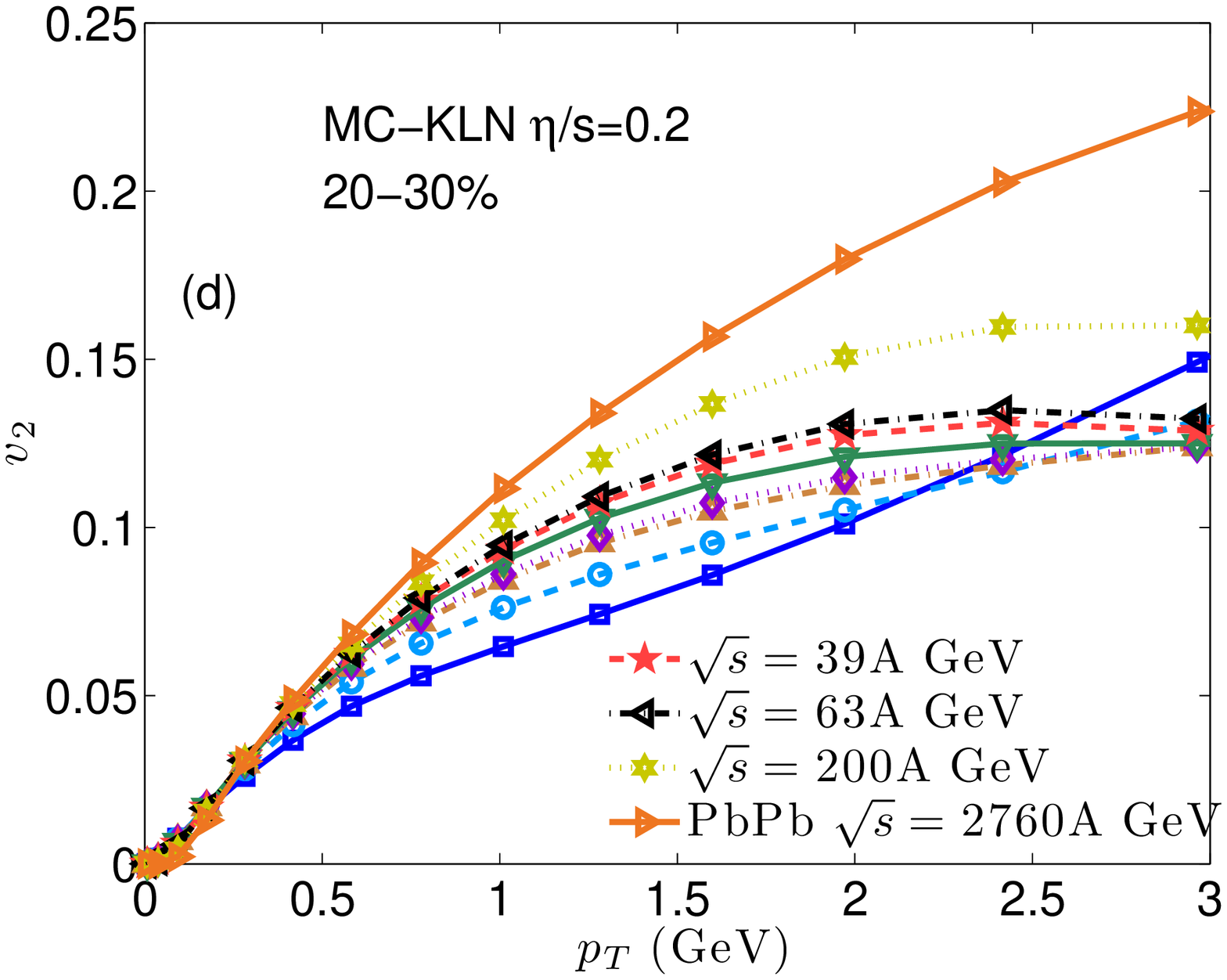}
  \end{tabular}
  \end{minipage}
  \begin{minipage}{0.2\linewidth}
  \caption{(a,c): Transverse momentum spectra of all charged hadrons from central 
  (0-5\% centrality) Au+Au and Pb+Pb collisions at 0-5\% centrality for different collision 
  energies. 
  (b,d): the corresponding differential elliptic flow at 20-30\% centrality.} 
  \label{fig3}
  \end{minipage}
\end{figure*}
%

A ``universal'' scaling behavior of the eccentricity-scaled elliptic flow as a function of charged hadron multiplicity density (``multiplicity scaling'') \cite{Alt:2003ab} was studied within viscous 
hydrodynamics in \cite{Song:2008si} and was later used to extract the specific shear viscosity from $\sqrt{s}=200$\,$A$\,GeV Au+Au collisions at RHIC \cite{Song:2010mg}. The authors of \cite{Hirano:2010jg} and \cite{Shen:2011eg} found that this ``universal'' scaling breaks down as $\sqrt{s}$ increases but disagreed on the sign of the scaling breaking effects. In Fig.~\ref{fig2} we explore the breaking of ``multiplicity scaling'' over a wider range of $\sqrt{s}$, for both of the initialization models. For MC-Glauber initial conditions (Fig.~\ref{fig2}a) eccentricity-scaled elliptic flow shows surprisingly good universality of the ``multiplicity scaling'' curve as the collision energy varies from 7.7 to 2760\,$A$\,GeV: The curves at different $\sqrt{s}$ fall almost perfectly on top of each other. For MC-KLN (Fig.~\ref{fig2}b), on the other hand, the ``universal scaling'' breaks in the same direction as previously shown in \cite{Shen:2011eg}: lower collision energies result in larger $v_2/\epsilon_2$ values at the same charged hadron multiplicity density. We found that the main reason for the different collision energy dependence between the MC-Glauber and MC-KLN models lies in the different centrality dependences of the initial overlap area in the two models. The initial overlap area is calculated as $S = \pi \sqrt{\langle x^2 \rangle \langle y^2 \rangle}$, where $\langle x^2 \rangle = \frac{\int d^2 \bm{r} \gamma e(\bm{r}) x^2 }{\int d^2 \bm{r} \gamma e(\bm{r})}$ is evaluated with the initial energy density as weight function.\footnote{The initial entropy density can also be used as weight. In \cite{Shen:2011eg} we showed that the scaling breaking behavior is independent of the choice of weight function.} As the collisions become more peripheral, the overlap area in the MC-KLN model decreases more rapidly than in the MC-Glauber model. In Pb+Pb collisions at $\sqrt{s}=2760$\,$A$\,GeV, the overlap area $S$ for MC-KLN decreases from 23.6\,fm$^2$ in the 0-5\% most central collisions to 4.7\,fm$^2$ in the 60-70\% centrality class; for MC-Glauber, $S$ decreases instead from 22.8\,fm$^2$ to 6.5\,fm$^2$. This slightly faster drop of the overlap area in the MC-KLN model shifts the ``universal'' scaling curves in Fig.~\ref{fig2} to the right and shrinks the covered range in $(1/S) dN_\mathrm{ch}/d\eta$. We further checked that the centrality dependence of the overlap area changes little as $\sqrt{s}$ varies from 7.7 to 2760\,$A$\,GeV. The different $\sqrt{s}$-dependences of $v_2/\epsilon_2$ as function of $dN_\mathrm{ch}/d\eta$ in Figs.~\ref{fig2}a and \ref{fig2}b thus reflect primarily the fact that the shape of the initial profiles evolves differently with centrality in the two initialization models. Fig.~\ref{fig2} can thus be used to check experimentally the consistency of the centrality dependence of the source size and shape in the initialization models.   

\begin{figure}[b]
  \includegraphics[width=0.86\linewidth]{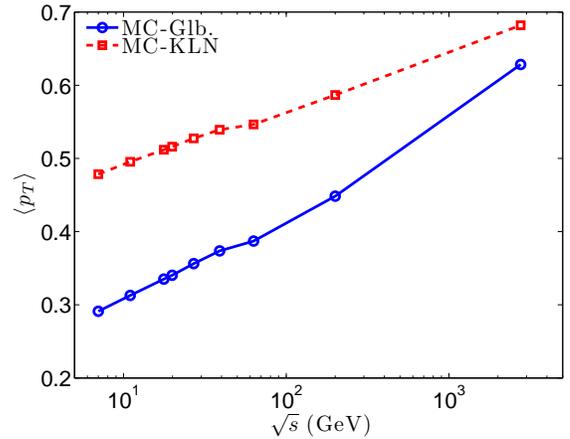} 
  \caption{Evolution with $\sqrt{s}$ of the average transverse momentum of 
  charged hadrons from central (0-5\% centrality) Au+Au and Pb+Pb collisions, for
  MC-Glauber and MC-KLN initial conditions.}
  \label{fig3'}
\end{figure}

\section{Charged particle $p_T$-spectra and differential elliptic flow}
\label{sec3}

\begin{figure*}
\begin{minipage}{0.79\linewidth}
  \begin{tabular}{cc}
  \includegraphics[width=0.48\linewidth]{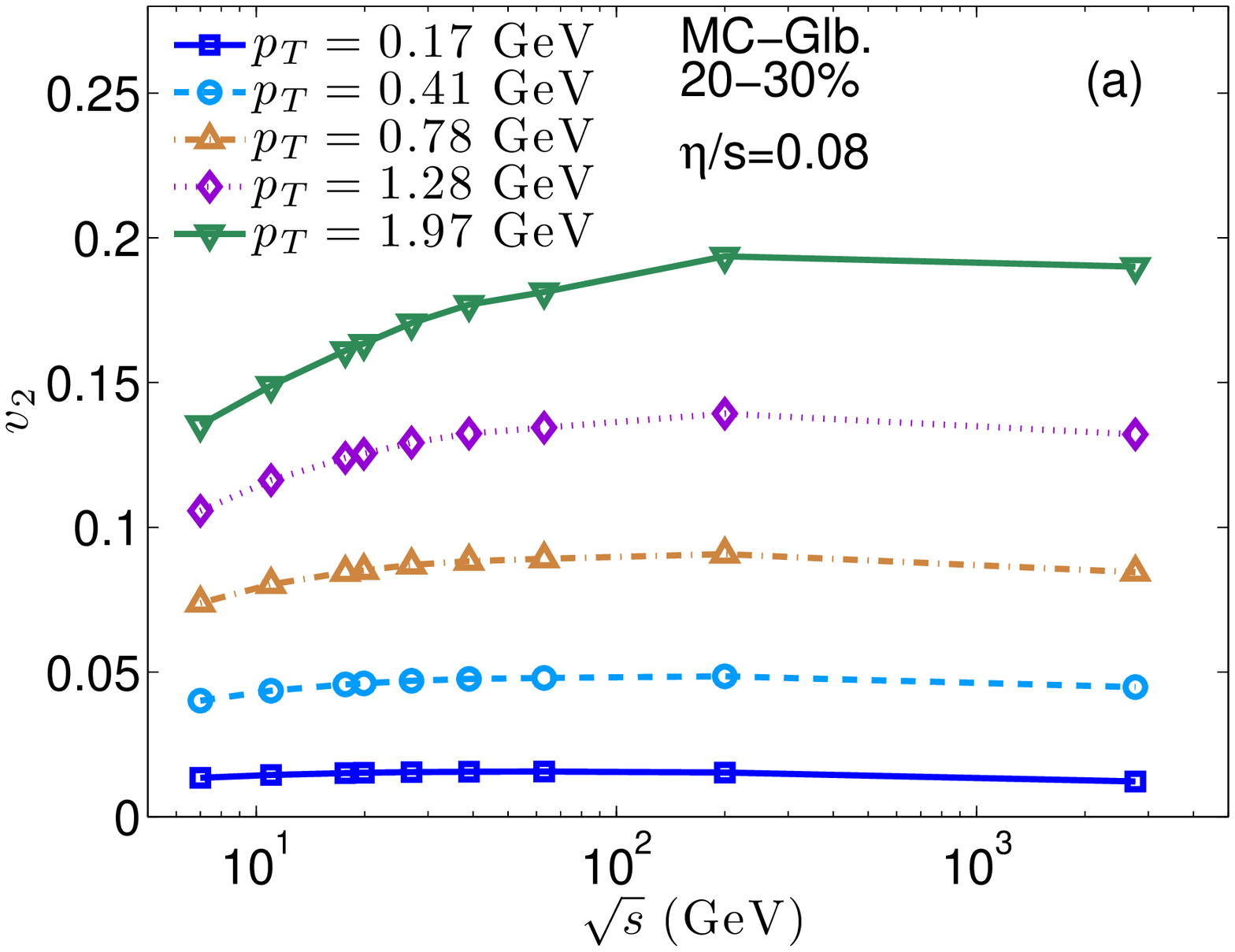} &
  \includegraphics[width=0.48\linewidth]{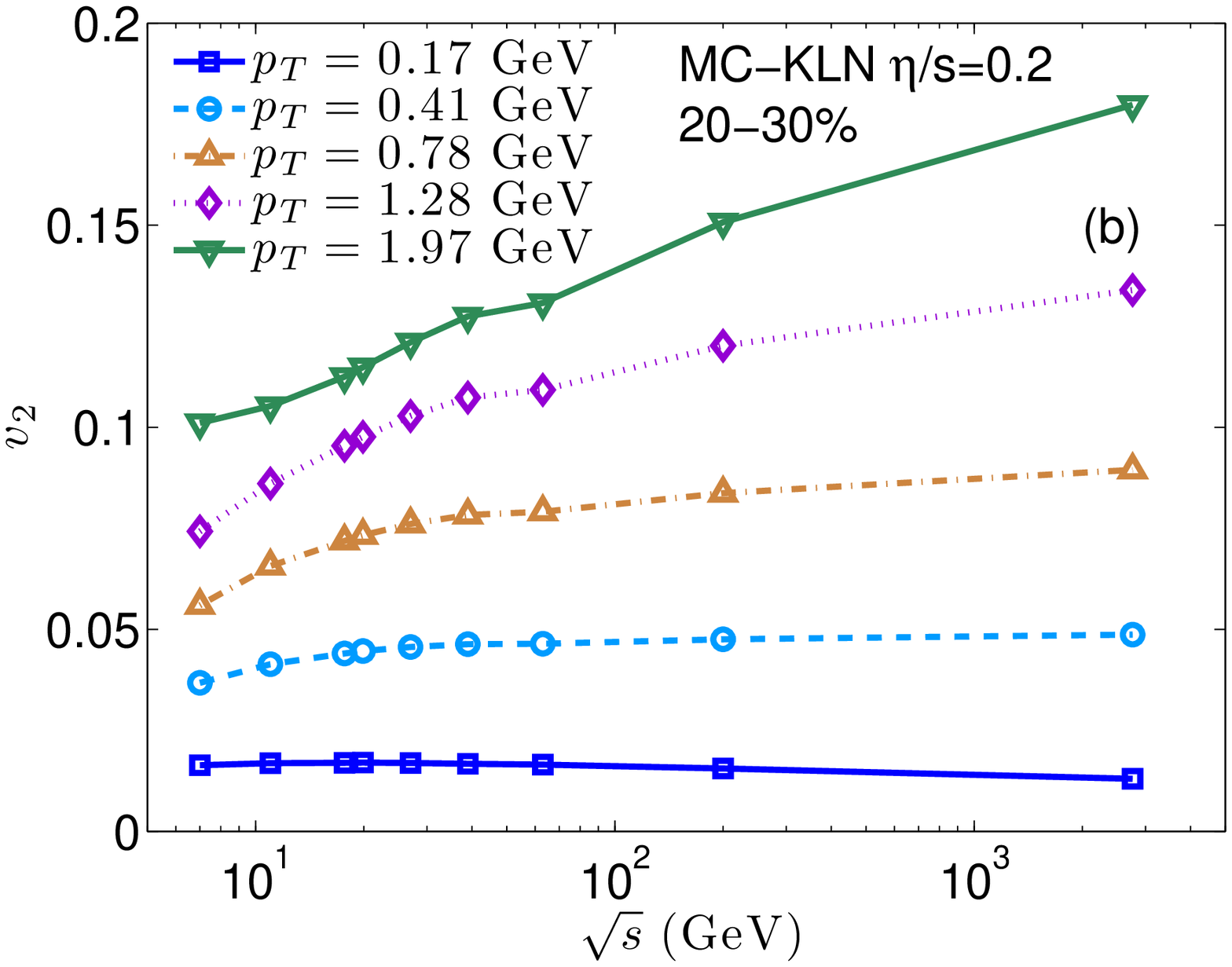}
  \end{tabular}  
\end{minipage}
\begin{minipage}{0.2\linewidth}
  \caption{Evolution with $\sqrt{s}$ of the differential charged hadron elliptic flow 
  $v_2^\mathrm{ch}(p_T,\sqrt{s})$ at 5 fixed $p_T$ values.}
  \label{fig4}
\end{minipage}
\end{figure*}

Figures~\ref{fig3}a,c show the $p_T$-spectra of all charged hadrons in the 0-5\% most central collisions. For both the MC-Glauber and MC-KLN models the slopes of the $p_T$-spectra get flatter as $\sqrt{s}$ increases: At higher collision energy the fireball lifetime is longer, which allows the system to develop more radial flow. The additional radial flow pushes more particles into the high-$p_T$ region, thus flattening the spectra. From $\sqrt{s} = 7.7$ to 2760\,$A$\,GeV, the mean $p_T$, $\langle p_T \rangle = \int d p_T p_T \frac{dN}{d\eta dp_T} / \int d p_T \frac{dN}{d\eta dp_T}$, increases by 43\% (from 0.48 to 0.68\,GeV/$c$) for the MC-KLN model and by 140\% (from 0.29 to 0.63\,GeV/$c$) for the MC-Glauber model (see Fig.~\ref{fig3'}).

The differential charged hadron elliptic flow is shown in Figs.\ref{fig3}b,d, for 20-30\% centrality. With MC-Glauber initial conditions the differential elliptic flow for $p_T < 2$ GeV remains almost unchanged for $\sqrt{s} \ge 39$\,$A$\,GeV. Below 39\,$A$\,GeV the slope of $v_2(p_T)$ begins to decrease. This tendency is indeed observed in the RHIC BES experiments \cite{Shi:2011ad,Pandit:2011hf,Schmah:2011zz}. We emphasize that our EOS s95p-PCE has no pronounced soft point in the phase transition region. This means that the often highlighted ``saturation'' of $v_2(p_T)$ above $\sqrt{s} = 39 A$\, GeV can not be associated with a softest point in the transition region. It is rather caused by a subtle cancellation of opposite $\sqrt{s}$-dependences of the differential $v_s(p_T)$ from light and heavy particles (see Fig.~\ref{fig7} below). 

For the MC-KLN model, the slope of the differential $v_2(p_T)$ decreases monotonically and continuously with decreasing collision energy. For a temperature-independent specific shear viscosity, $\eta/s=0.2$, the collision energy dependence of the differential elliptic flow observed here is somewhat inconsistent with the experimental observation of a $v_2^\mathrm{ch}(p_T)$ that does not change between $\sqrt{s}=39$ and 2760$A$\,GeV. Within the MC-KLN framework, this might be taken as an indication for a possible temperature dependence of $\eta/s$ \cite{Niemi:2011ix,Shen:2011kn,Shen:2011eg}. Additional studies are, however, necessary to fully address this issue \cite{Heinz:2011kt}. For $\sqrt{s} = 7.7$ and 11.5\,$A$\,GeV, the differential $v_2$ is seen to increase more quickly above $p_T > 2.5$\,GeV. We find this to be caused by large $\delta f$ corrections (i.e. non-equilibrium corrections arising from non-zero shear stresses at freeze out \cite{Shen:2011eg}). The larger $\delta f$ corrections at lower collision energies indicate a narrowing of the temporal interval during which viscous hydrodynamics is a valid description. At lower $p_T$ ($p_T < 2$\,GeV), our results show monotonic $\sqrt{s}$ dependence. 

To further illustrate this point we plot in Fig.~\ref{fig4} the $\sqrt{s}$-dependence of $v_2^\mathrm{ch}(p_T)$ at 5 fixed $p_T$ points. In this representation one sees that for the MC-Glauber model $v_2^\mathrm{ch}$ at any fixed $p_T$-value features as a function of $\sqrt{s}$ a very broad maximum somewhere around top RHIC energy (200\,$A$\,GeV); for low $p_T<0.5$\,GeV/$c$, this maximum occurs at lower $\sqrt{s}$. A similar behavior was seen in \cite{Kestin:2008bh} for
ideal hydrodynamics with a bag-model equation of state which features a strong minimum (``softest point") in the speed of sound at the phase transition temperature. We see that the existence of this maximum does not depend on the appearance of a softest point in the EOS.
Compared to the earlier ideal fluid calculations, the position where $v_2^\mathrm{ch}$ at fixed $p_T$ assumes its largest value has been shifted to larger $\sqrt{s}$ values by viscous effects. This shift is seen to be even stronger in the MC-KLN case (Fig.~\ref{fig4}b) where the fluid is much more viscous. For shear viscosities as large as those needed to describe the $v_2^\mathrm{ch}$ measured in 200\,$A$\,GeV Au+Au collisions with MC-KLN initial conditions ($\eta/s{\,\simeq\,}2.5/(4\pi)$) \cite{Luzum:2009sb,Song:2010mg}, $v_2^\mathrm{ch}(p_T,\sqrt{s})$ at fixed $p_T$
has not yet reached its maximum value even at top LHC energies (except for very small $p_T{\,<\,}200$\,MeV/$c$).  

\begin{figure*}
  \begin{minipage}{0.79\linewidth}
  \begin{tabular}{cc}
  \includegraphics[width=0.49\linewidth,height=5.8cm]{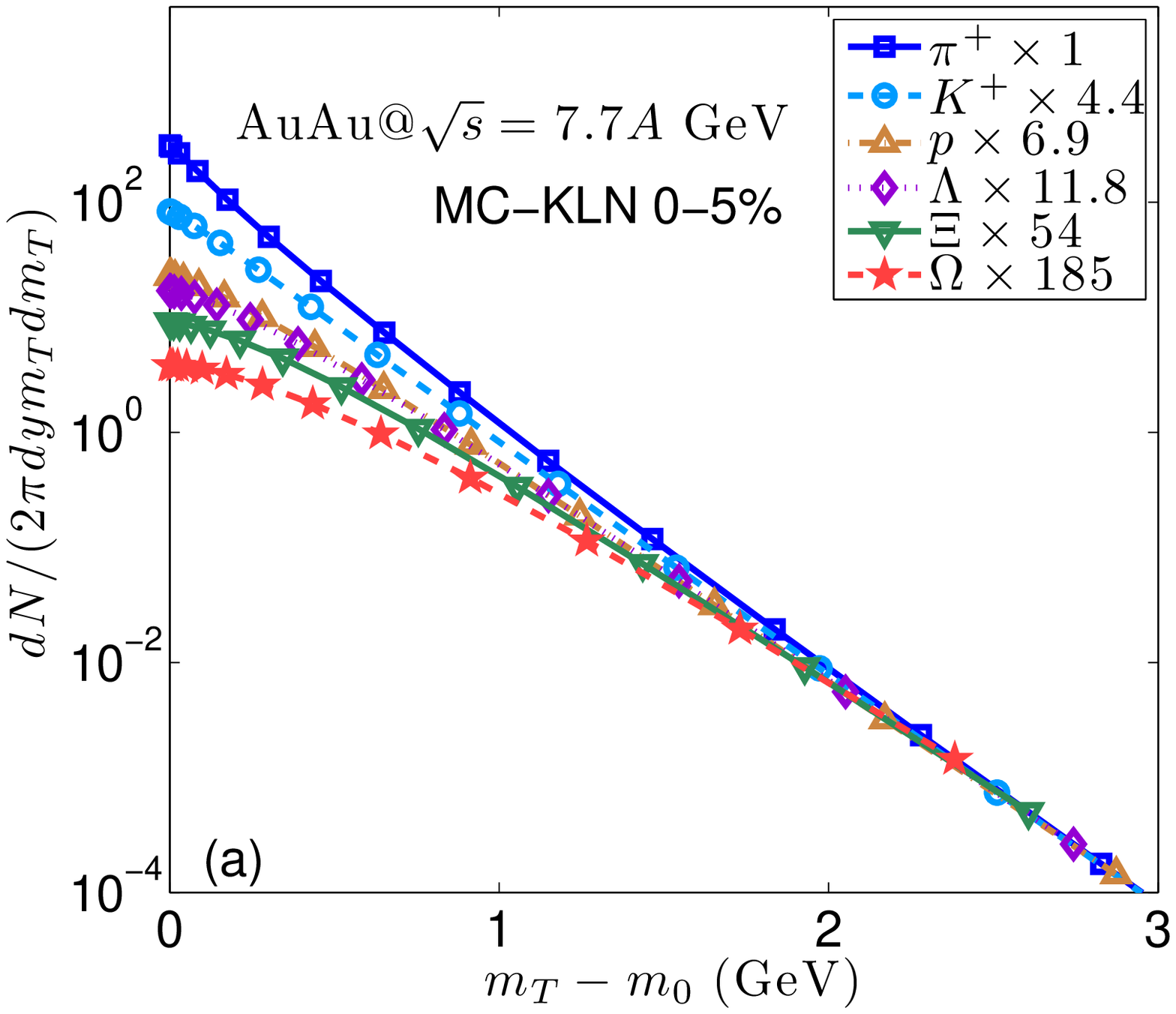} &
  \includegraphics[width=0.49\linewidth,height=5.8cm]{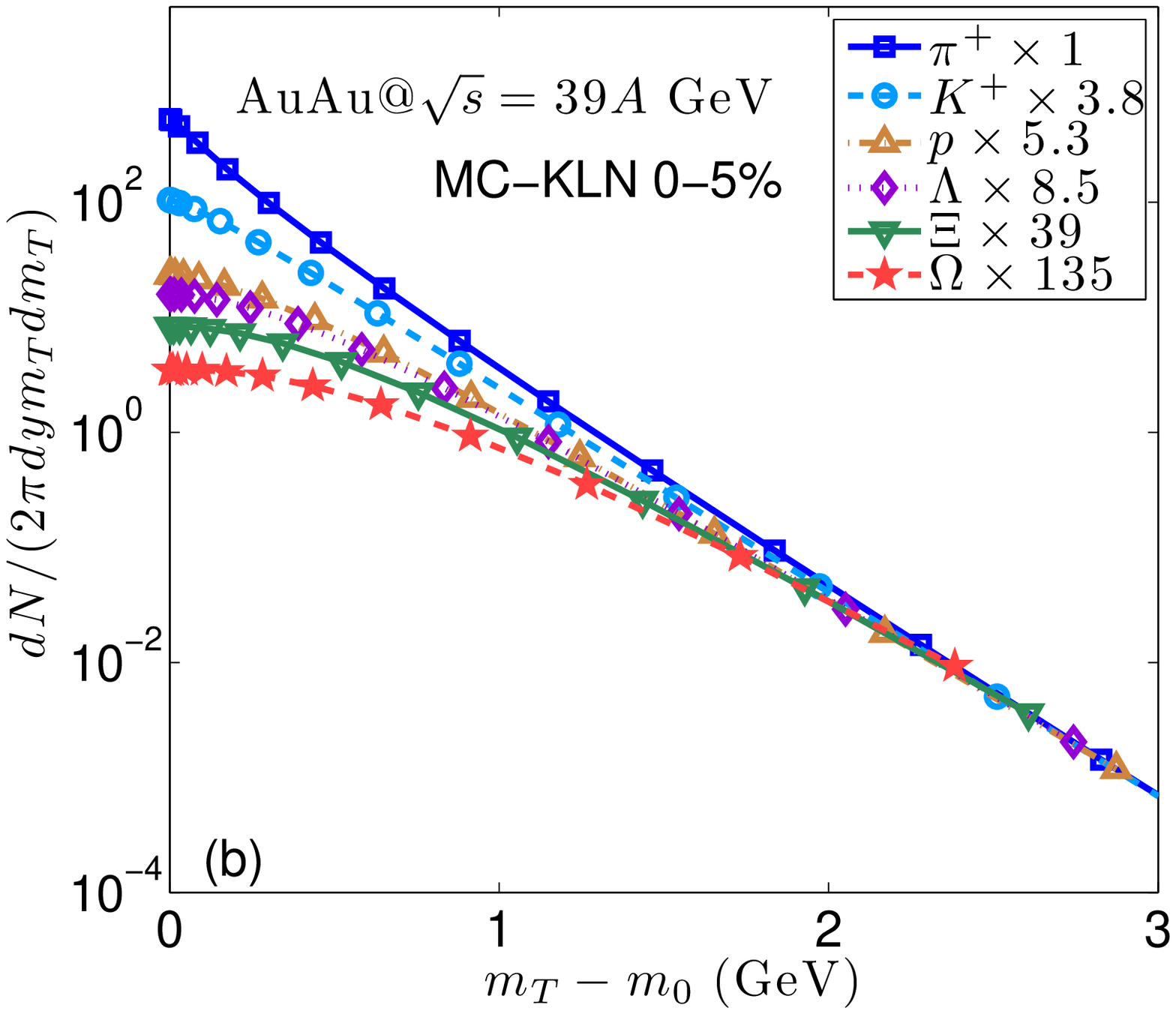} \\
  \includegraphics[width=0.49\linewidth,height=5.8cm]{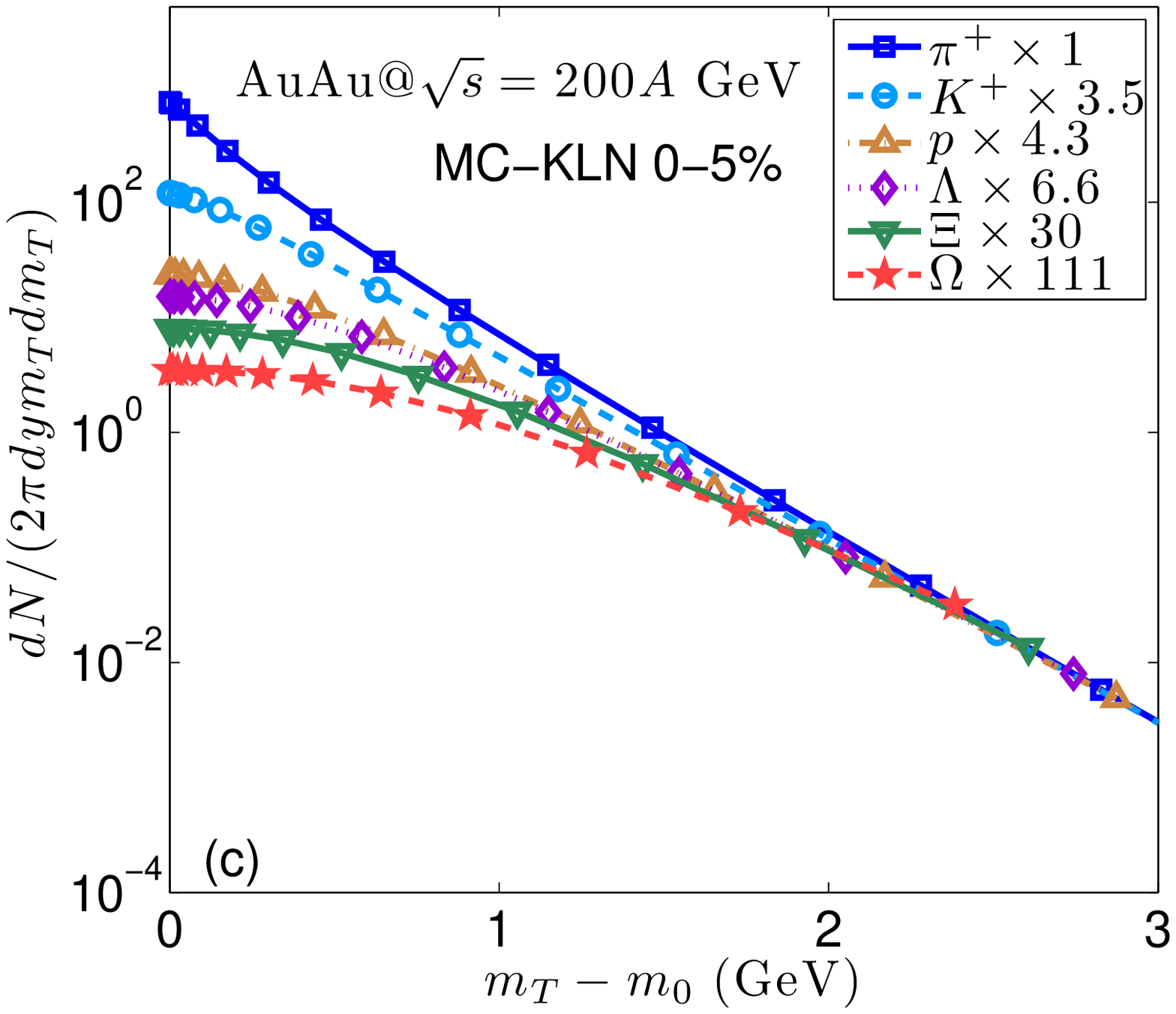} &
  \includegraphics[width=0.49\linewidth,height=5.8cm]{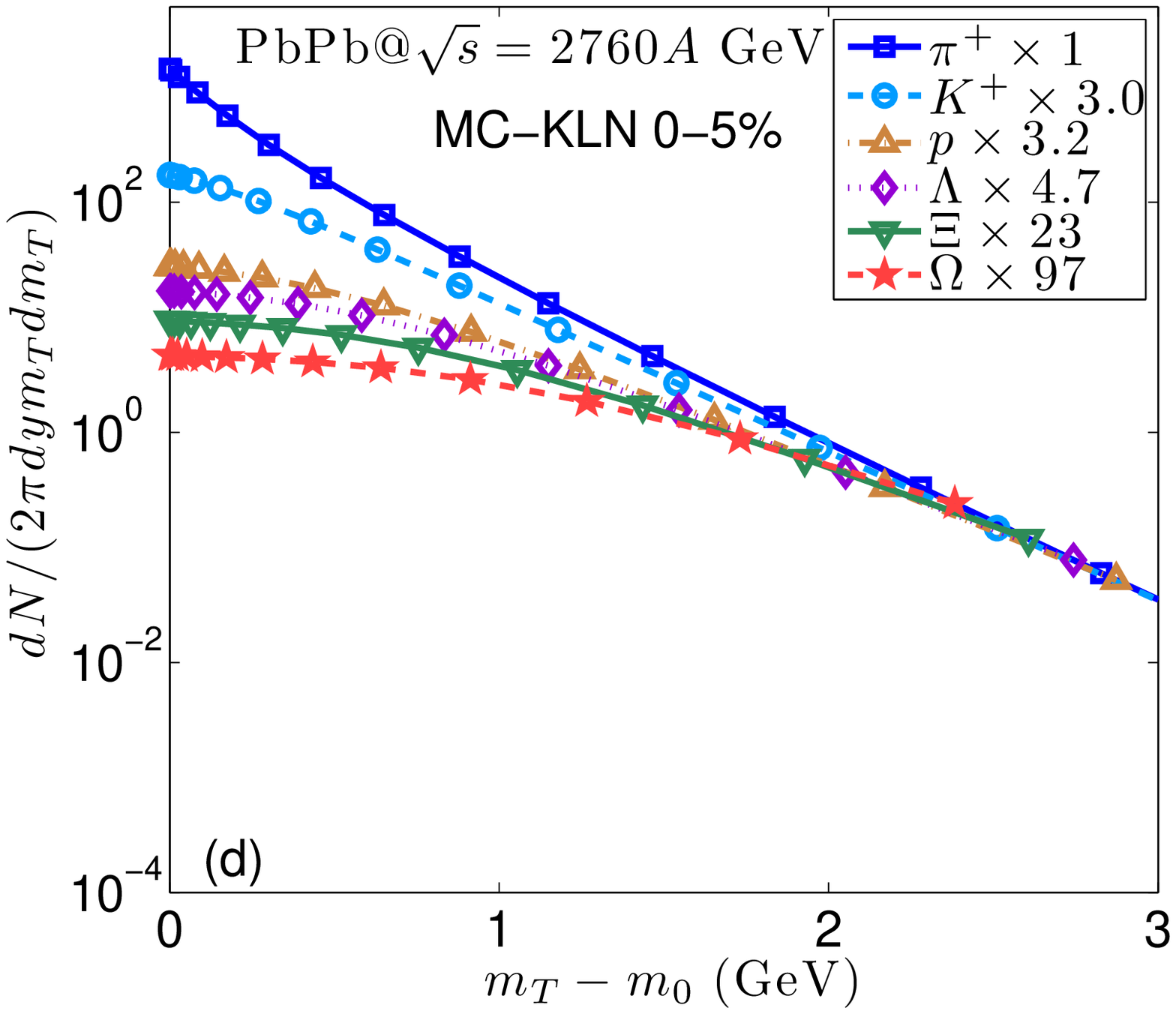} 
  \end{tabular}
 \end{minipage}
 \begin{minipage}{0.2\linewidth}
   \caption{Identified particle spectra as a function of $m_T{-}m_0$ for the MC-KLN model 
   in the 0-5\% most central collisions, at $\sqrt{s}{\,=\,}7.7,\ 39,\ 200$, and $2760$\,$A$\,GeV. 
   The spectra for MC-Glauber initial conditions look qualitatively similar.}
  \label{fig5}
 \end{minipage}
\end{figure*}

At the lower end of the $\sqrt{s}$-range studied in Fig.~\ref{fig4}, the increase with collision energy of $v_2(p_T,\sqrt{s})$ at fixed $p_T$ is a consequence of increasing fireball lifetimes which allow the initial spatial eccentricity of the fireball to convert more fully into anisotropic hydrodynamic flow. At higher collision energies eventually the point is reached where this momentum anisotropy is fully saturated before the system falls apart; longer fireball lifetimes will then no longer lead to more anisotropic flow, only to more radial flow. Stronger radial flow, however,
pushes the momentum anisotropy out to larger $p_T$, by generating flatter $p_T$ distributions. As a result, elliptic flow at fixed $p_T$ begins to decrease. In practice, this radial flow driven decrease of $v_2(p_T)$ at fixed $p_T$ sets in even before the $p_T$-integrated total charged hadron elliptic flow $v_2^\mathrm{ch}$ has reached saturation \cite{Heinz:2009ny}, and it accelerates thereafter. 

\section{$p_T$-spectra and elliptic flow of identified hadrons}
\label{sec4}

We now proceed to study how hydrodynamical flow affects identified particles. 

It is well known that thermal spectra from a static fireball exhibit $m_T$-scaling, $dN_i / (2\pi dy m_T dm_T) \sim \sqrt{m_T}\,e^{-m_T/T}$ \cite{Heinz:2004qz}, and that radial flow breaks this scaling. In order to isolate the radial flow effects we therefore plot in Fig.~\ref{fig5} the $m_T$-spectra of identified particles as a function of $m_T{-}m_0$, for four selected $\sqrt{s}$ values. Except for minor effects from the viscous $\delta f$ corrections, resonance feed-down and Bose statistics for pions, in the absence of flow the slopes of the $m_T$-spectra would be the same for all hadron species. To show the flow-induced slope difference, we scaled in Figs.~\ref{fig5} the heavy particle spectra by constant factors to the same value at $m_T{-}m_0{\,=\,}3$\,GeV. At large $m_T{-}m_0$ rest mass effects become negligible, and all hadrons have approximately the same inverse slope $T_\mathrm{eff} = T_\mathrm{dec} \sqrt{(1{+}\langle v_\perp \rangle)/(1{-}\langle v_\perp \rangle)}$ \cite{Heinz:2004qz}. In Fig.~\ref{fig5} we find that for low $\sqrt{s}$ values $m_T$-scaling is significantly broken only at $m_T{-}m_0 < 2$\,GeV, while at LHC energy the flow-induced breaking of $m_T$-scaling extends to 3\,GeV of transverse kinetic energy. At low $m_T{-}m_0$ the spectra are split by hadron mass effects, and this splitting increases with $\sqrt{s}$ due to the increasing radial flow which pushes heavier particles to larger $p_T$. At the highest collision energy $\sqrt{s}=2760$\,$A$\,GeV we observe a particularly strong concavity of the pion spectra at low $m_T{-}m_0$, due to Bose statistics. For other mesons, their heavy rest masses suppress Bose effects. 

\begin{figure*}
  \begin{minipage}{0.79\linewidth}
  \begin{tabular}{cc}
  \includegraphics[width=0.49\linewidth,height=5.8cm]{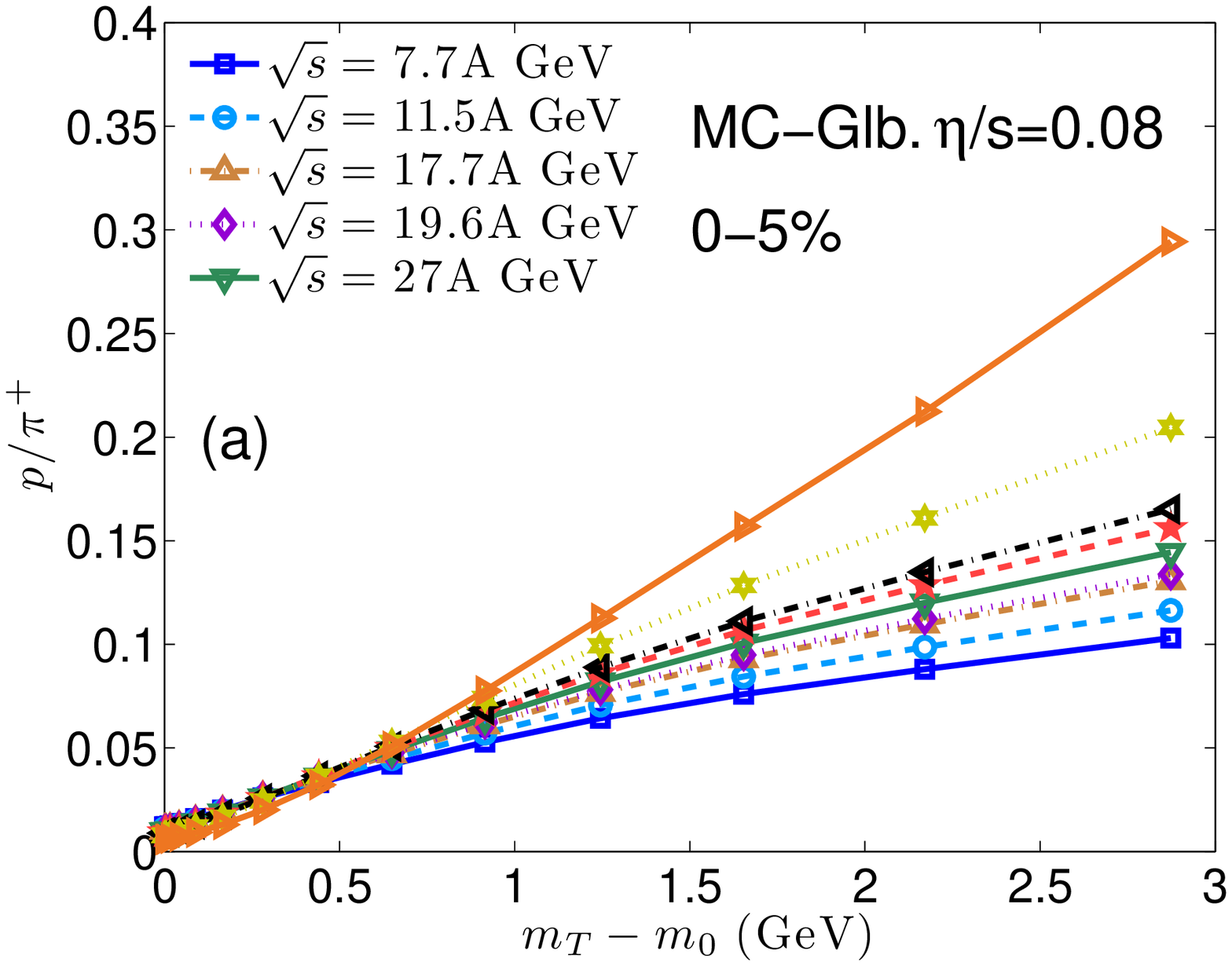} &
  \includegraphics[width=0.49\linewidth,height=5.8cm]{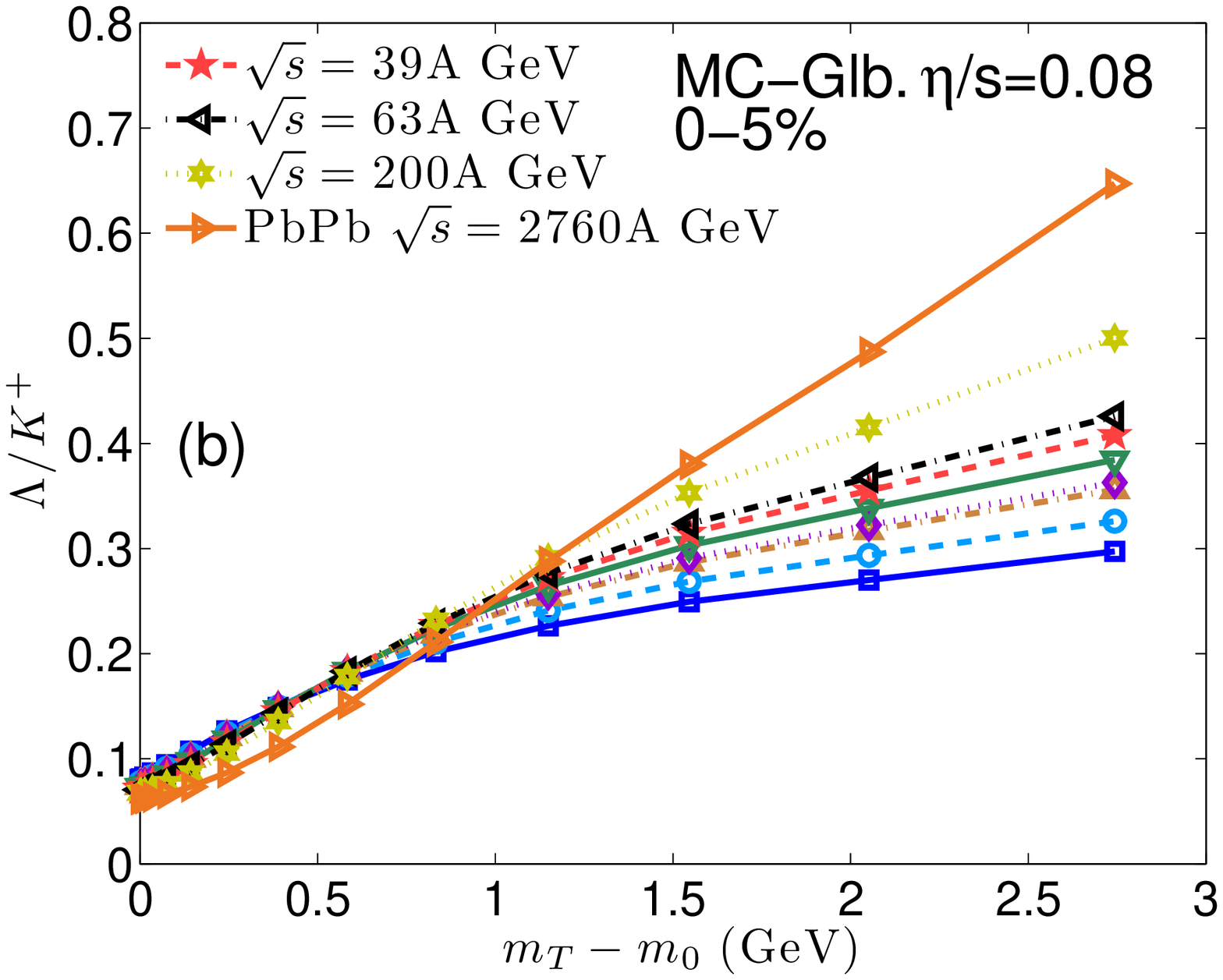} \\
  \includegraphics[width=0.49\linewidth,height=5.8cm]{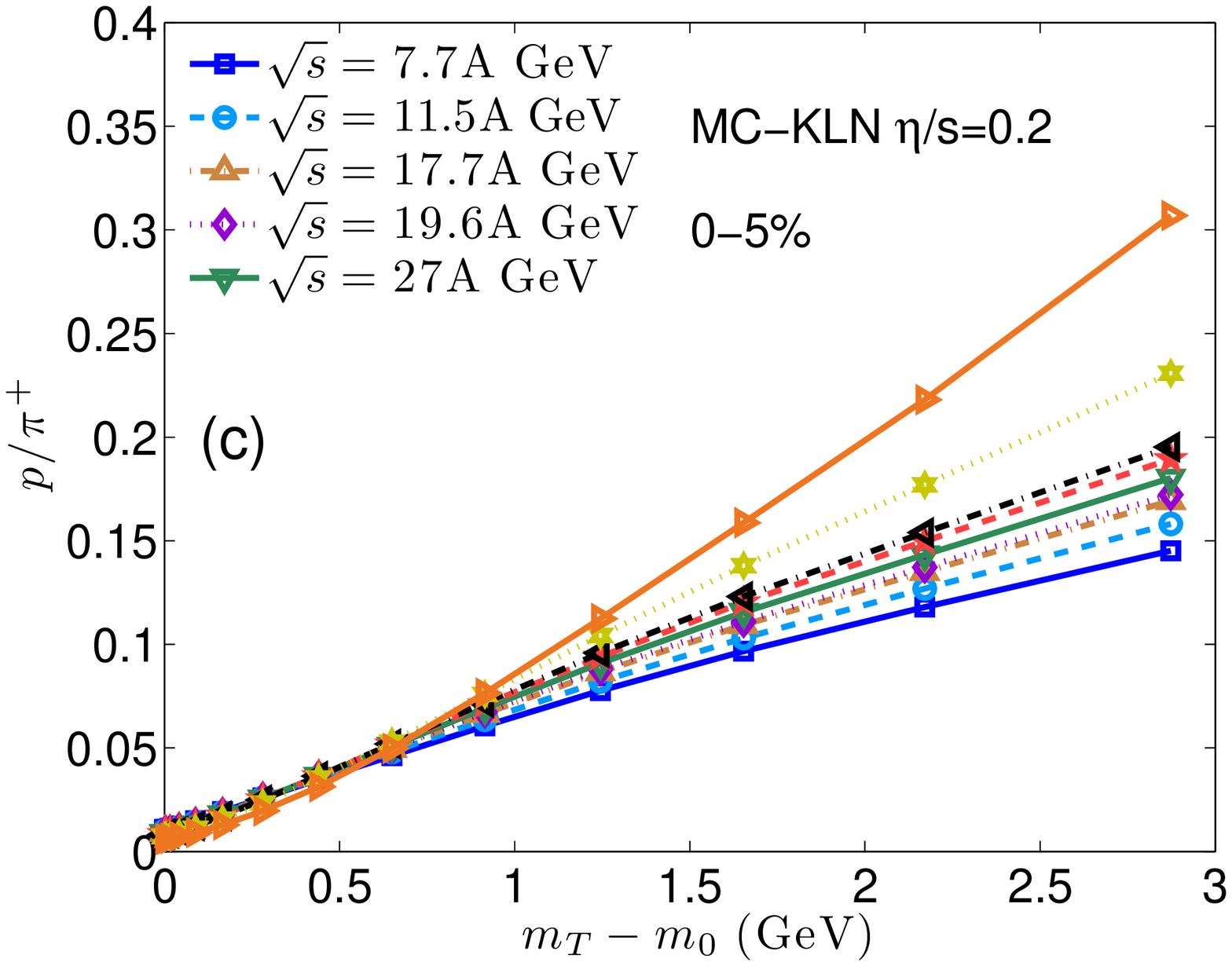} &
  \includegraphics[width=0.49\linewidth,height=5.8cm]{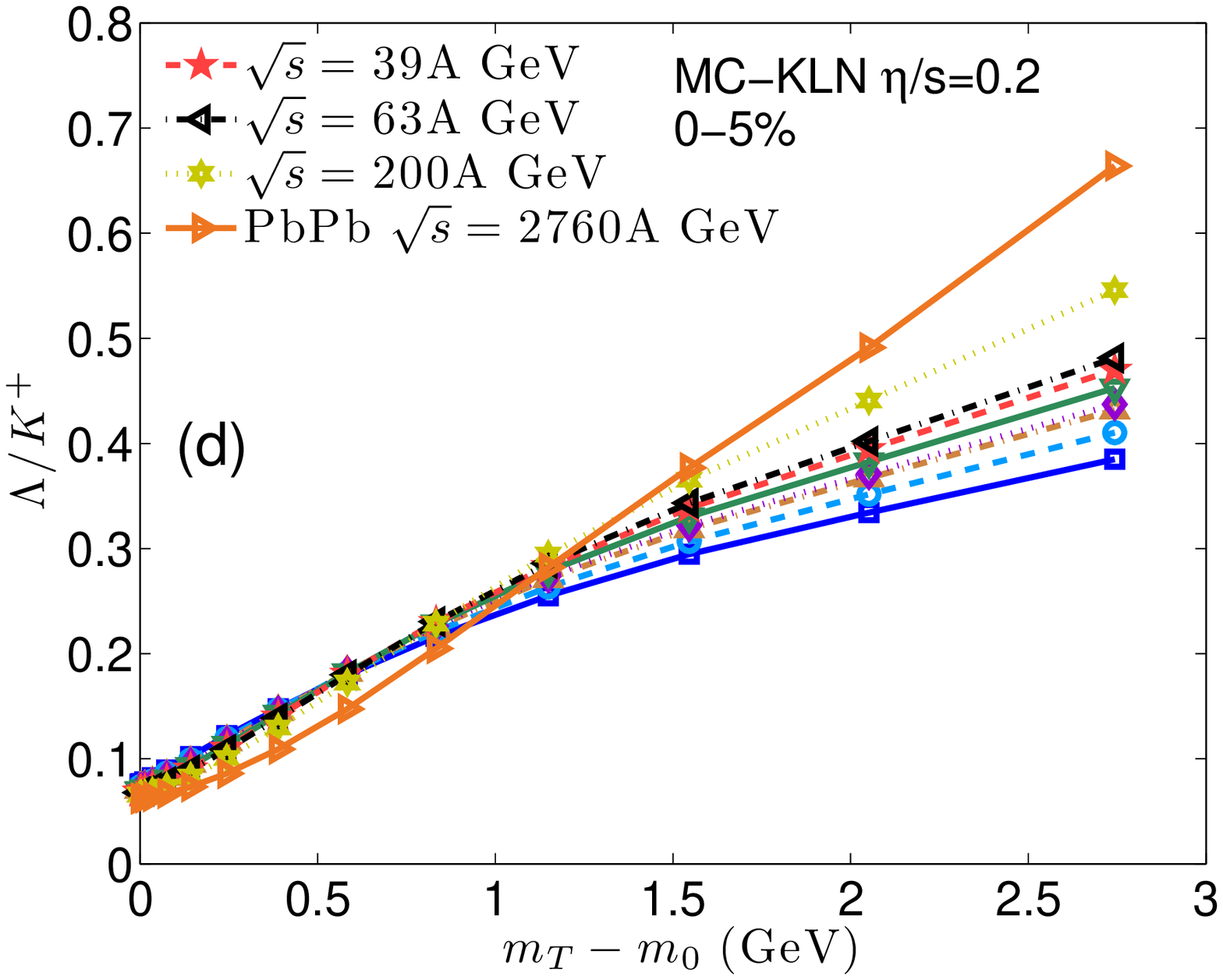}
  \end{tabular}
  \end{minipage}
  \begin{minipage}{0.2\linewidth}
  \caption{The $p/\pi^+$ (a,c) and $\Lambda/K^+$ (b,d) ratios as functions of $m_T{-}m_0$, 
  for MC-Glauber (a,b) and MC-KLN (c,d) initial conditions at 0-5\% most central collisions, 
  from $\sqrt{s}=7.7$ to 2760\,$A$\,GeV. Please note the dramatic increase of radial flow 
  effects on these ratios between RHIC and LHC energies.}
  \label{fig6}
  \end{minipage}
\end{figure*}

The flow-induced breaking of $m_T$-scaling is seen even more clearly when one plots heavy-to-light particle ratios (such as $p/\pi^+$, $\Lambda/K^+$) as a function of transverse kinetic energy. For a static thermalized fireball, these ratios should be independent of $m_T{-}m_0$, up to small quantum statistical corrections arising from  the pion spectra at small $m_T{-}m_0$. Fig.~\ref{fig6} shows that for an expanding fireball these ratios increase with increasing transverse kinetic energy, at a rate that itself increases with $\sqrt{s}$, reflecting the larger radial flow at higher collision energies. A little more careful inspection and thought reveal that, in fact, stronger radial flow {\it increases} the $p/\pi^+$ and $\Lambda/K^+$ ratios at large $m_T{-}m_0$ while {\it decreasing} them at small $m_T{-}m_0$. This is so because in our simulations the $p_T$-integrated particle ratios are the same at all collision energies, as we assumed zero baryon chemical potential and the same chemical and kinetic freeze-out temperatures at all $\sqrt{s}$. In addition, radial flow flattens the $m_T$-dependence of these ratios at low $m_T{-}m_0$, due to the ``flow shoulder'' developing in the heavy-particle $m_T$-spectra at low transverse kinetic energy when radial flow gets strong. This shoulder is weaker for protons than for $\Lambda$'s, but in the $p/\pi^+$ ratio the more prominent Bose effect in the pion spectra at high collision energies additionally helps to flatten out the $p/\pi^+$ ratio at small $m_T{-}m_0$. Overall, Fig.~\ref{fig6} shows that these features are all very similar for MC-Glauber and MC-KLN initial conditions.

\begin{figure*}
  \begin{minipage}{0.79\linewidth}
  \begin{tabular}{cc}
  \includegraphics[width=0.49\linewidth,height=5.8cm]{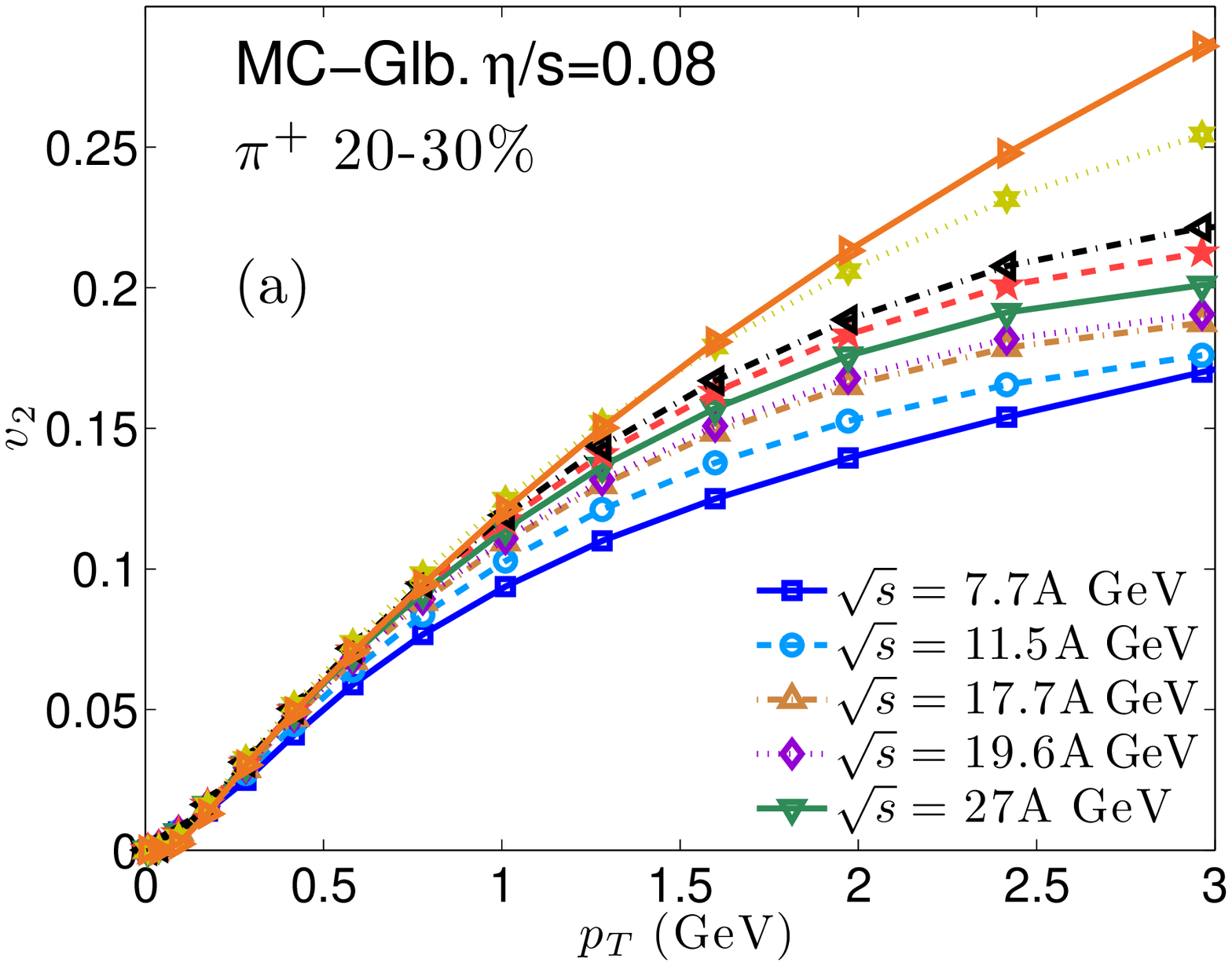} &
  \includegraphics[width=0.49\linewidth,height=5.8cm]{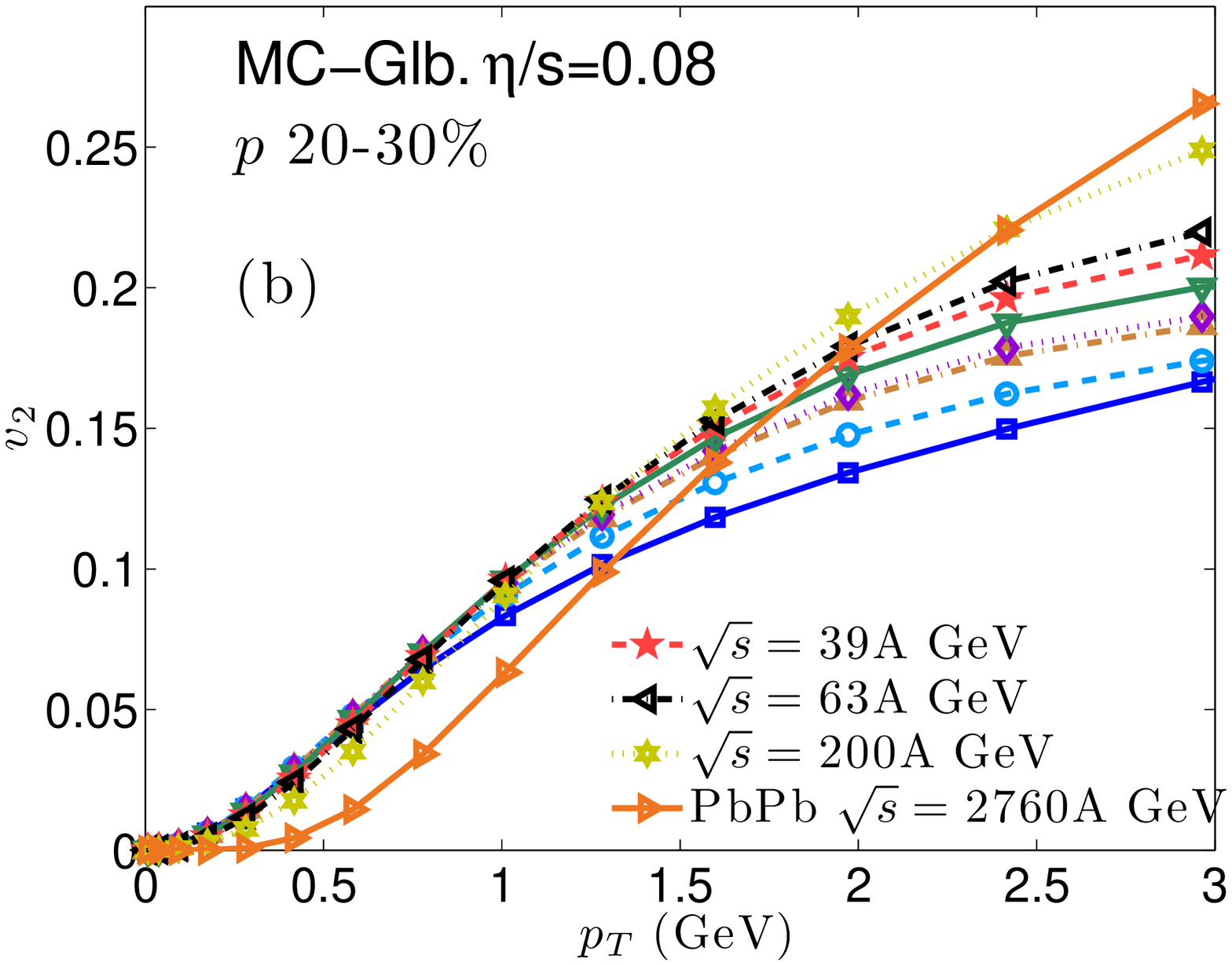} \\
  \includegraphics[width=0.49\linewidth,height=5.8cm]{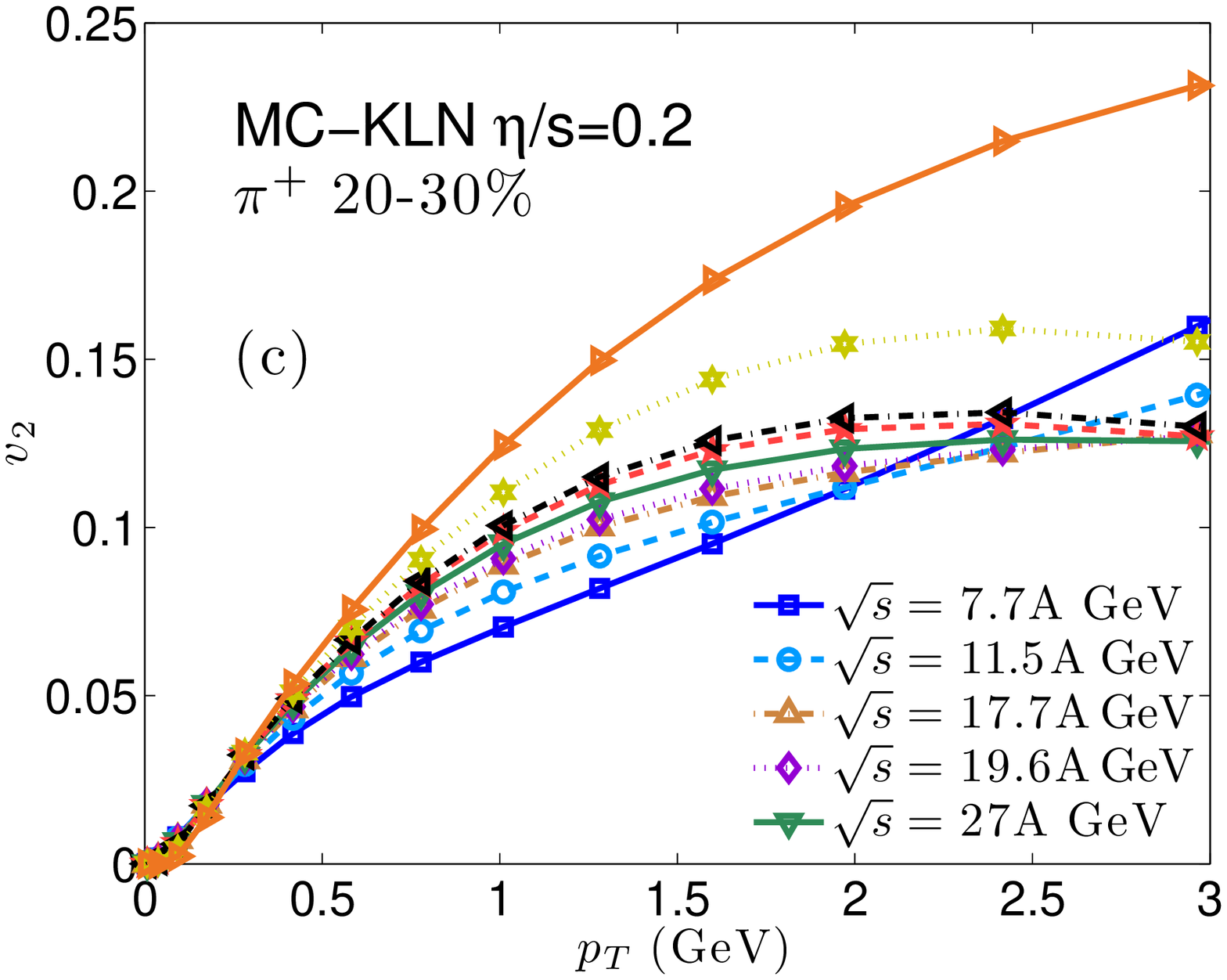} &
  \includegraphics[width=0.49\linewidth,height=5.8cm]{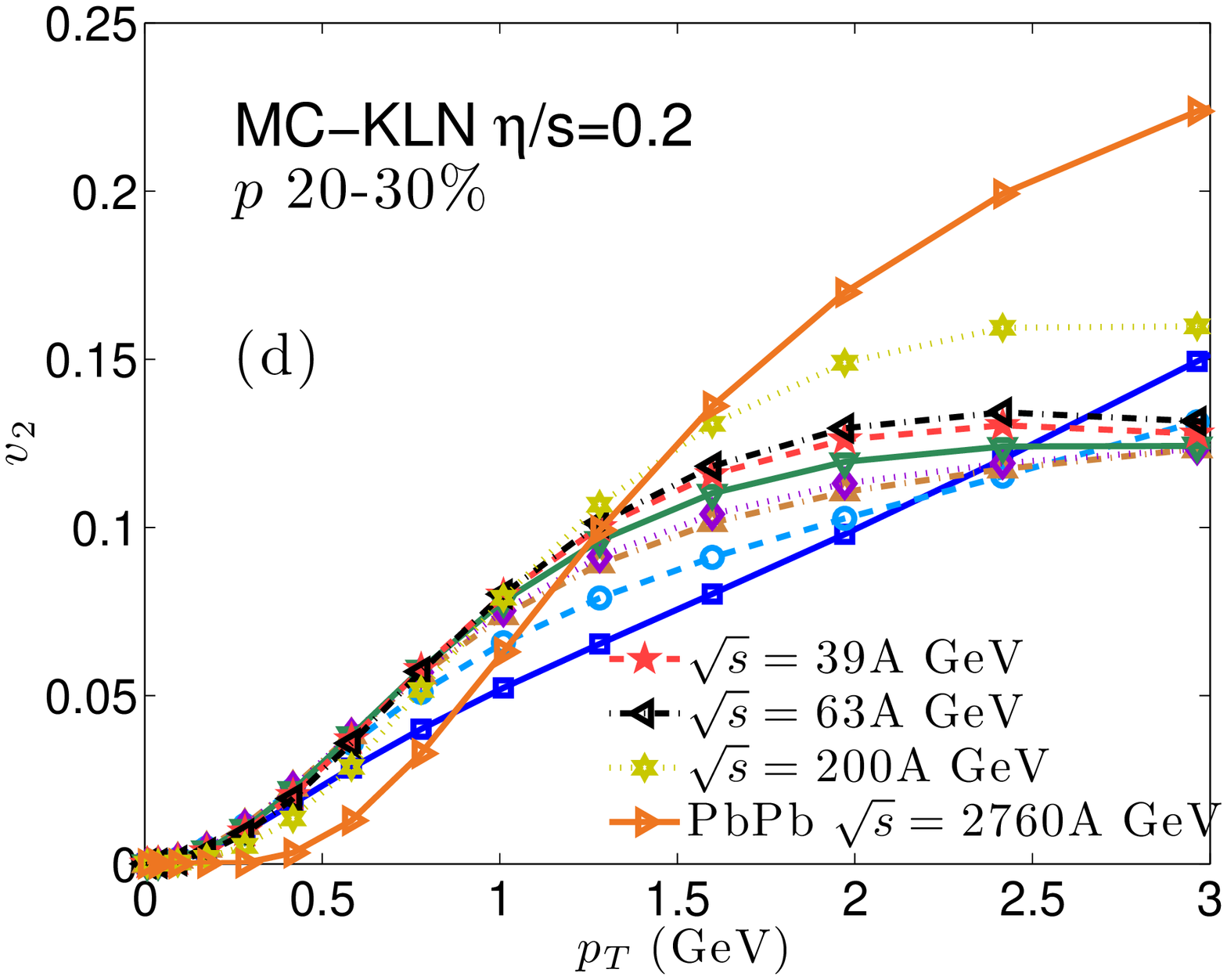}
  \end{tabular}
  \end{minipage}
  \begin{minipage}{0.2\linewidth}
  \caption{Differential elliptic flow of $\pi^+$ (a,c) and $p$ (b,d) at 20-30\% centrality, 
  for MC-Glauber (a,b) and MC-KLN (c,d) profiles.}
  \label{fig7}
  \end{minipage}
\end{figure*}

In Fig. \ref{fig7} we show the differential elliptic flow of $\pi^+$ and $p$ for $\sqrt{s}=7.7$ to $2760$\,$A$\,GeV. For pions, the differential $v_2$ varies with $\sqrt{s}$ very similarly to the total charged hadron elliptic flow shown in Figs.~\ref{fig3}b,d. For protons the strong radial flow ``blueshifts'' the entire elliptic flow to higher $p_T$. So for higher collision energies, the values of $v_2$ are smaller in the low $p_T$-region and larger in the high-$p_T$ region. We find that below 200\,$A$\,GeV the proton $v_2(p_T)$ at low $p_T$ is almost independent of $\sqrt{s}$. At LHC energy, on the other hand, the blueshift is really dramatic, reflecting the much stronger radial flow at this high collision energy. The total charged hadron elliptic flow is the combination of contributions from light pions and less abundant heavy particles. Since with increasing collision energy the elliptic flow of heavy particles decreases at low $p_T$, they effectively cancel the weak increase of the light pion $v_2$. This results in the apparent saturation of charged hadron differential elliptic flow over a wide $p_T$ range from $\sqrt{s}=39$-$2760$\,$A$\,GeV that was seen in Fig.~\ref{fig3}b for MC-Glauber model. For the MC-KLN model, this cancellation is less efficient because the increase with $\sqrt{s}$ of the pion $v_2$ is stronger (see Fig.~\ref{fig7}c,d). Therefore, for the MC-KLN model the total charged hadron elliptic flow keeps increasing as the collision energy increases.

\section{Spatial eccentricity at freeze-out}
\label{sec5}

We conclude this paper by presenting a novel shape analysis of the evolving fireball. Theoretically, the spatial eccentricity $\epsilon_x$ is conventionally defined at fixed proper time $\tau$ by
\begin{equation}
\epsilon_x (\tau) = \frac{\int dx\,dy\, (y^2{-}x^2) \gamma e(x,y;\tau)}
                                     {\int dx\,dy\, (y^2{+}x^2) \gamma e(x,y;\tau)}.
\label{eq4}
\end{equation}
The weight function $\gamma e(x,y;\tau)$ is the energy density in the laboratory frame.\footnote{The entropy density $s(x,y;\tau)$ can also be used as weight function for calculating the eccentricity. The authors of  \cite{Qiu:2011iv} concluded from a study of fluctuating initial conditions that these two definitions yield initial spatial eccentricities that are linearly related to each other although the actual values are slightly different.}

Since the measured hadrons are only emitted from the final kinetic freeze-out surface, experimentalists can only infer the shape of that surface, by exploiting two-particle momentum correlations among the emitted particles and their dependence on the azimuthal angle around the beam axis \cite{Lisa:2000ip,Retiere:2003kf}. For comparison with such experimentally determined final source eccentricities \cite{Barrette:1997fj,Lisa:2000hw,Adams:2004yc,Abelev:2009tp}, a more meaningful theoretical quantity would be the spatial eccentricity of the final freeze-out surface $\Sigma$,
\begin{equation}
\epsilon_x (\Sigma) = \frac{\int_\Sigma u^\mu d^3 \sigma_\mu \,(y^2{-}x^2)}
                                          {\int_\Sigma u^\mu d^3 \sigma_\mu\,(y^2{+}x^2) },
\label{eq5}
\end{equation}
instead of a constant proper time surface. In Eq.~(\ref{eq5}) $u^{\mu}$ is the flow velocity on the surface $\Sigma$, and we used the fact that for our EOS an isothermal freeze-out surface is also a surface of constant local energy or entropy density, and that therefore the weight functions $e(x, y, \tau)$ or $s(x, y, \tau)$ cancel between numerator and denominator.\footnote{If we do not cancel the weight function and replace $\Sigma$ by a constant proper time surface, the definition (\ref{eq5}) reduces to Eq.~(\ref{eq4}) for longitudinally boost-invariant systems.} Thus the spatial eccentricity (\ref{eq5}) defined on an isothermal surface is independent of whether we weight $\epsilon_x$ with energy or entropy density.  

\begin{figure}[b]
  \includegraphics[width=0.95\linewidth]{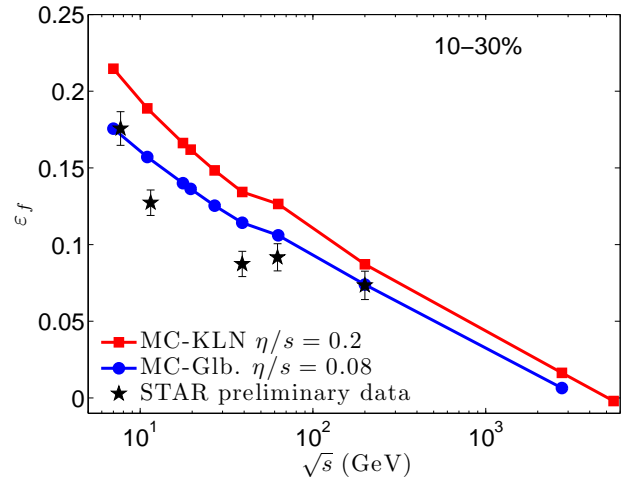}
  \caption{$\sqrt{s}$-dependence of the final spatial eccentricity $\varepsilon_\mathrm{f}$ of 
  the isothermal kinetic freeze-out surface at $T_\mathrm{dec}{\,=\,}120$\,MeV, for 10-30\% 
  centrality. The initial eccentricity is 0.26 for the MC-Glauber model and 0.32 for the MC-KLN
  model. The experimental points indicate preliminary data \cite{Anson:2011ik} from an 
  azimuthal HBT analysis by the STAR Collaboration.}
  \label{fig8}
\end{figure}

In Fig.~\ref{fig8}, we show the final eccentricity calculated along the kinetic freeze-out surface, $T_\mathrm{dec}=120$\,MeV, as a function of collision energy. For both MC-Glauber and MC-KLN models, as the collision energy increases, the final spatial eccentricity $\varepsilon_\mathrm{f}$ decreases monotonically. This is because at higher collision energy the system lives longer, giving the fireball more time to decompress and (due to anisotropic flow) become less deformed. For sufficiently large initial energy density, the fireball has actually enough time to become elongated along the reaction plane, instead of its original elongation perpendicular to it \cite{Heinz:2002sq,Frodermann:2007ab}. In Fig. \ref{fig8} we compare our results with recent (preliminary) STAR data from an azimuthal HBT analysis \cite{Anson:2011ik}. MC-Glauber runs with $\eta/s{\,=\,}0.08$ quantitatively reproduce the data at $\sqrt{s}{\,=\,}200A$\,GeV while underpredicting the final eccentricity by $\sim$10\% at lower energies. MC-KLN initial conditions with $\eta/s{\,=\,}0.2$ result in 15-20\% larger final eccentricities than both the MC-Glauber runs and the STAR data, due to the $\sim$20\% larger initial eccentricities of the MC-KLN profiles. In Table~\ref{table1} we see that, at the same $\sqrt{s}$, the fireball lifetimes with MC-KLN and MC-Glauber initial conditions are very similar. In spite of the faster evolution of radial flow for the MC-KLN initial conditions, the larger initial eccentricity in the MC-KLN model is preserved all the way to the end of the hydrodynamic evolution. Extending our calculations to LHC energy we predict that $\varepsilon_\mathrm{f}$ will approach zero around $\sqrt{s} = 2.76$-5.5\,$A$\,TeV. Again, this is a result of the longer fireball lifetime at LHC energies. For even larger $\sqrt{s}$, $\varepsilon_\mathrm{f}$ will turn negative, in qualitative agreement with previous calculations in \cite{Heinz:2002sq,Frodermann:2007ab} using ideal fluid dynamics and a less realistic EOS. Contrary to our present work, the authors of \cite{Heinz:2002sq,Frodermann:2007ab} calculated the azimuthal HBT radii from the Cooper-Frye output of their hydrodynamic simulations. In future work we will calibrate our definition (\ref{eq5}) for $\varepsilon_\mathrm{f}$ against the final eccentricity value extracted from azimuthal oscillations of HBT radii. Here we only note that an azimuthal HBT analysis at LHC energy will help to further test predictions from the viscous hydrodynamic model. 

The careful reader may have noticed that there is a small kink at $\sqrt{s} = 63$\,$A$\,GeV in the slopes of the solid and dashed lines in Fig. \ref{fig8} and earlier in Figs.~\ref{fig3'} and \ref{fig4}. Unfortunately, this is not a phase transition signature but rather an artifact from our normalization of the initial energy density profile to experimentally measured final charged multiplicity data for $\sqrt{s} \ge 63$\,$A$\,GeV and to the empirical formula Eq.~(\ref{eq2}) for $\sqrt{s} \le 39$\,$A$\,GeV. At $\sqrt{s}=63$\,$A$\,GeV the experimentally measured $dN_\mathrm{ch}/d\eta$ is $\sim5\%$ smaller than the value obtained from Eq.~(\ref{eq2}). If we use Eq.~(\ref{eq2}) instead of the measured value to normalize our initial energy density profiles at $\sqrt{s}=63$\,$A$\,GeV, the fireball lifetime increases slightly, decreasing the final eccentricity by a few precent and removing the kink in the theoretical curves. The origin of an apparently similar kink in the STAR data which happens to occur (we believe: accidentally) also around $\sqrt{s} = 63 A$\,GeV deserves further study. 
 
\section{Summary}
\label{sec6}

Using (2+1)-dimensional viscous hydrodynamics with properly implemented hadronic chemical freeze-out at $T_\mathrm{chem}{\,=\,}165$\,MeV, we have studied systematically the evolution of hydrodynamic observables with collision energy in the range $7.7{\,\leq\,}\sqrt{s}{\,\leq\,}2760\,A$\,GeV. Over this range of energies, the initial peak temperature almost doubles and the fireball lifetime increases by about 60\%. We find that for temperature independent specific shear viscosity the MC-Glauber model shows almost perfect ``multiplicity-scaling" of the eccentricity-scaled charged hadron elliptic flow. For the MC-KLN model this scaling is broken: as $\sqrt{s}$ increases, the $v_2^\mathrm{ch}/\varepsilon_2$ vs. $(1/S)(dN_\mathrm{ch}/dy)$ curves shift to the right (Fig.~\ref{fig2}b). We found that this breaking of multiplicity scaling in the MC-KLN model originates from a steeper centrality dependence of the nuclear overlap area.

For both initialization models, higher collision energies generate stronger radial flow which results in flatter hadron spectra and a corresponding increase of the mean $p_T$ of charged hadrons. For the MC-Glauber model we observed an approximate ``saturation" of the charged hadron differential elliptic flow at fixed $p_T$ in the region $\sqrt{s}{\,\geq\,}39$\,$A$\,GeV, similar to what is observed experimentally. We believe, however, that the word ``saturation" describes the observations incorrectly and that what is seen is better described as a very broad maximum (as a function of $\sqrt{s}$) of the differential elliptic flow at fixed $p_T$, caused by the interplay of (i) growing total momentum anisotropy (which increases $v_2$) and increasing radial flow (which decreases $v_2$ at fixed $p_T$ by shifting it to larger $p_T$), and (ii) increasing $v_2(p_T)$ for pions and decreasing $v_2(p_T)$ for kaons, protons and other heavy hadrons between RHIC and LHC energies. The mechanism (i) causes maxima of $v_2(p_T)$ at fixed $p_T$ for all hadron species, but located at lower $\sqrt{s}$ values for heavier than for lighter hadrons (due to the mass-dependence of radial flow effects on the $p_T$-spectra). The mechanism (ii) ensures that the maximum for all charged hadrons is broader in $\sqrt{s}$ than for each hadron species individually and thus manifests itself as a broad plateau that (for $\eta/s{\,=\,}0.08$) happens to span the collision energy range from upper RHIC to LHC energies. The position in $\sqrt{s}$ of the maximum of $v_2(p_T)$ at fixed $p_T$ for each hadron species depends on the viscosity of the fluid (which controls the interplay in the development of radial and elliptic flow during the fireball expansion) and increases with $\eta/s$. For MC-KLN initial conditions, which require $\sim$2.5 times larger $\eta/s$ for a successful description of elliptic flow data at RHIC and LHC, $v_2^\mathrm{ch}(p_T)$ at fixed $p_T$ has not yet reached its maximal value even at LHC energies. 

Finally, we have proposed an improved measure for the final fireball eccentricity at kinetic freeze-out and studied its evolution with collision energy. It is found to decrease monotonically with increasing collision energy, at a rate that is roughly consistent with recent experimental measurements. Its absolute value agrees with the data better for the MC-Glauber than for the MC-KLN model -- the $\sim$20\% larger initial eccentricities of the MC-KLN profiles yield final freeze-out eccentricities that again appear to be $\sim$20\% larger than those from MC-Glauber initial profiles, and lie significantly above the measured values. Neither model describes the available data perfectly; in view of the limitations of the purely hydrodynamic approach employed here ({\it cf.} our discussion in the Introduction) this is not too surprising. The model predicts, however, robustly that at top RHIC energies the final freeze-out source is still out-of-plane elongated (as experimentally observed), but that at LHC energies the final eccentricity should approach zero. Measurements that test this prediction should soon become available.  


\acknowledgments{We thank Christopher Anson and Mike Lisa for stimu\-lating discussions.This work was supported by the U.S. Department of Energy under Grants No. DE-SC0004286 and (within the framework of the JET Collaboration) DE-SC0004104.}



\begin{thebibliography}{99}

\bibitem{Kumar:2011de}
  L.~Kumar {\it et al.} (STAR Collaboration),
  Nucl.\ Phys.\  {\bf A862-A863}, 125 (2011).
  
\bibitem{Caines:2009yu} 
  H.~Caines {\it et al.} (STAR Collaboration),
  arXiv:0906.0305 [nucl-ex].
  
\bibitem{Shi:2011ad} 
  S.~Shi {\it et al.} (STAR Collaboration),
  arXiv:1111.5385 [nucl-ex].

\bibitem{Pandit:2011hf}
  Y.~Pandit {\it et al.} (STAR Collaboration),
  J.\ Phys.\ Conf.\ Ser.\  {\bf 316}, 012001 (2011)
  
\bibitem{Kumar:2011us} 
  L.~Kumar {\it et al.} (STAR Collaboration),
  J.\ Phys.\ G {\bf 38}, 124145 (2011)
  
\bibitem{Schmah:2011zz} 
  A.~Schmah (STAR Collaboration),
  J.\ Phys.\ G {\bf 38}, 124049 (2011).
  
\bibitem{Alford:2007xm} 
  M.~G.~Alford, A.~Schmitt, K.~Rajagopal and T.~Sch\"afer,
  Rev.\ Mod.\ Phys.\  {\bf 80}, 1455 (2008).
  
\bibitem{Fodor:2001pe} 
  Z.~Fodor and S.~D.~Katz,
  JHEP {\bf 0203}, 014 (2002).
  
\bibitem{Ejiri:2003dc} 
  S.~Ejiri, C.~R.~Allton, S.~J.~Hands, O.~Kaczmarek, F.~Karsch, E.~Laermann and C.~Schmidt,
  Prog.\ Theor.\ Phys.\ Suppl.\  {\bf 153}, 118 (2004).
  
\bibitem{Gavai:2004sd} 
  R.~V.~Gavai and S.~Gupta,
  Phys.\ Rev.\ D {\bf 71}, 114014 (2005).
  
\bibitem{Teaney:2003kp}
  D.~Teaney,
  Phys.\ Rev.\  C {\bf 68}, 034913 (2003).
  
\bibitem{Lacey:2006pn}
  R.~A.~Lacey and A.~Taranenko,
  PoS {\bf CFRNC2006}, 021 (2006);
  R.~A.~Lacey {\it et al.},
  Phys.\ Rev.\ Lett.\  {\bf 98}, 092301 (2007);
  A.~Adare {\it et al.},
  Phys.\ Rev.\ Lett.\  {\bf 98}, 172301 (2007);
  H.-J.~Drescher, A.~Dumitru, C.~Gombeaud, and J.-Y. Ollitrault,
  Phys.\ Rev.\  C {\bf 76}, 024905 (2007).
  K.~Dusling and D.~Teaney,
  Phys.\ Rev.\  C {\bf 77}, 034905 (2008);
  Z.~Xu, C.~Greiner, and H.~St\"ocker,
  Phys.\ Rev.\ Lett.\  {\bf 101}, 082302 (2008);
  D.~Molnar and P.~Huovinen,
  J.\ Phys.\ G {\bf 35}, 104125 (2008);
  R.~A.~Lacey, A.~Taranenko and R.~Wei,
  in {\it Proc. 25th Winter Workshop on Nuclear Dynamics},
  W. Bauer, R. Bellwied, and J.W. Harris (eds.),
  (EP Systema, Budapest, 2009) p. 73 [arXiv:0905.4368 [nucl-ex]];
  K.~Dusling, G.~D.~Moore, and D.~Teaney,
  Phys.\ Rev.\  C {\bf 81}, 034907 (2010);
  A.~K.~Chaudhuri,
  J.\ Phys.\ G {\bf 37}, 075011 (2010);
  R.~A.~Lacey {\it et al.},
  Phys.\ Rev.\  C {\bf 82}, 034910 (2010).

\bibitem{Romatschke:2007mq}
  P.~Romatschke and U.~Romatschke,
  Phys.\ Rev.\ Lett.\  {\bf 99}, 172301 (2007).

\bibitem{Song:2007fn} 
  H.~Song and U.~Heinz,
  Phys.\ Lett.\ B {\bf 658}, 279 (2008),
  and Phys.\ Rev.\ C {\bf 77}, 064901 (2008).

\bibitem{Song:2008si}
  H.~Song and U.~Heinz,
  Phys.\ Rev.\ C {\bf 78}, 024902 (2008).

\bibitem{Luzum:2008cw}
  M.~Luzum and P.~Romatschke,
  Phys.\ Rev.\  C {\bf 78}, 034915 (2008).

\bibitem{Luzum:2009sb}
  M.~Luzum and P.~Romatschke,
  Phys.\ Rev.\ Lett.\  {\bf 103}, 262302 (2009).

\bibitem{Song:2010mg}
  H.~Song, S.~A.~Bass, U.~Heinz, T.~Hirano and C.~Shen,
  Phys.\ Rev.\ Lett.\ {\bf 106}, 192301 (2011);
  and Phys.\ Rev.\ C {\bf 83}, 054910 (2011).

\bibitem{Aamodt:2010pa}
  K.~Aamodt {\it et al.} (ALICE Collaboration),
  Phys.\ Rev.\ Lett.\  {\bf 105}, 252302 (2011).

\bibitem{Luzum:2010ag}
  M.~Luzum,
  Phys.\ Rev.\  C {\bf 83}, 044911 (2011).

\bibitem{Lacey:2010ej}
  R.~A.~Lacey, A.~Taranenko, N.~N.~Ajitanand and J.~M.~Alexander,
  Phys.\ Rev.\  C {\bf 83}, 031901 (2011).

\bibitem{Bozek:2010wt}
  P.~Bozek, M.~Chojnacki, W.~Florkowski and B.~Tomasik,
  Phys.\ Lett.\ {\bf B694}, 238 (2010);
  P.~Bozek,
  {\it ibid.} {\bf B699}, 283 (2011).

\bibitem{Hirano:2010jg}
  T.~Hirano, P.~Huovinen and Y.~Nara,
  Phys.\ Rev.\  C {\bf 83}, 021902 (2011).

\bibitem{Schenke:2010rr}
  B.~Schenke, S.~Jeon and C.~Gale,
  Phys.\ Rev.\ Lett.\  {\bf 106}, 042301 (2011).
  
\bibitem{Schenke:2011tv}
  B.~Schenke, S.~Jeon and C.~Gale,
  Phys.\ Lett.\  {\bf B702}, 59 (2011).

\bibitem{Song:2011qa}
  H.~Song, S.~A.~Bass and U.~Heinz,
  Phys.\ Rev.\ C {\bf 83}, 054912 (2011).

\bibitem{Shen:2010uy}
  C.~Shen, U.~Heinz, P.~Huovinen and H.~Song,
  Phys.\ Rev.\ C {\bf 82}, 054904 (2010).

\bibitem{Shen:2011eg}
  C.~Shen, U.~Heinz, P.~Huovinen and H.~Song,
  Phys.\ Rev.\ C {\bf 84}, 044903 (2011).

\bibitem{Alver:2010gr}
  B.~Alver and G.~Roland,
  Phys.\ Rev.\  C {\bf 81}, 054905 (2010).

\bibitem{Adare:2011tg} 
  A.~Adare {\it et al.}  (PHENIX Collaboration),
  Phys.\ Rev.\ Lett.\  {\bf 107}, 252301 (2011);
  R.~Lacey {\it et al} (PHENIX Collaboration),
  J.\ Phys.\ G {\bf 38}, 124048 (2011).

\bibitem{Sorensen:2011fb}
  P.~Sorensen{\it et al.} (STAR Collaboration),
  J.\ Phys.\ G {\bf 38}, 124029 (2011)

\bibitem{ALICE:2011vk}
  K. Aamodt {\it et al.} (ALICE Collaboration),
  Phys.\ Rev.\ Lett.\  {\bf 107}, 032301 (2011);
  R.~Snellings {\it et al.} (ALICE Collaboration),
  J.\ Phys.\ G {\bf 38}, 124013 (2011);
  M.~Krzewicki {\it et al.} (ALICE Collaboration), 
  J.\ Phys.\ G {\bf 38}, 124047 (2011).

\bibitem{CMSflow}
  S. Chatrchyan {\it et al.} (CMS Collaboration),
  CERN preprint CMS-PAS-HIN-11-005;
  %
  J.~Velkovska {\it et al.} (CMS Collaboration),
  J.\ Phys.\ G {\bf 38}, 124011 (2011).
    
\bibitem{Steinberg:2011dj}
  P.~Steinberg {\it et al.} (ATLAS Collaboration),
  J.\ Phys.\ G {\bf 38}, 124004 (2011)
  J.~Jia {\it et al.}  (ATLAS Collaboration),
  J.\ Phys.\ G {\bf 38}, 124012 (2011)

\bibitem{Alver:2010dn}
  B.~H.~Alver, C.~Gombeaud, M.~Luzum, and J.~Y.~Ollitrault,
  Phys.\ Rev.\  C {\bf 82}, 034913 (2010).

\bibitem{Petersen:2010cw}
  H.~Petersen, G.-Y.~Qin, S.~A.~Bass and B.~M\"uller,
  Phys.\ Rev.\  C {\bf 82}, 041901 (2010).

\bibitem{Qin:2010pf}
  G.-Y.~Qin, H.~Petersen, S.~A.~Bass and B.~M\"uller,
  Phys.\ Rev.\  C {\bf 82}, 064903 (2010).
  
\bibitem{Luzum:2010sp}
  M.~Luzum,
  Phys.\ Lett.\  {\bf B696}, 499 (2011).
  
\bibitem{Xu:2011fe}
  J.~Xu and C.~M.~Ko,
  Phys.\ Rev.\  {\bf C84}, 014903 (2011).

\bibitem{Luzum:2011mm}
  M.~Luzum,
   J.\ Phys.\ G {\bf 38}, 124026 (2011). 

\bibitem{Qiu:2011hf} 
  Z.~Qiu, C.~Shen and U.~Heinz,
  Phys.\ Lett.\ B {\bf 707}, 151 (2012).
  
\bibitem{Chaudhuri:2011qm}
  A.~K.~Chaudhuri,
  arXiv:1108.5552 [nucl-th].

\bibitem{Schenke:2011bn} 
  B.~Schenke, S.~Jeon and C.~Gale,
  Phys.\ Rev.\ C {\bf 85}, 024901 (2012).
  
\bibitem{Lacey:2010hw}
  R.~A.~Lacey, R.~Wei, J.~Jia, N.~N.~Ajitanand, J.~M.~Alexander, and A.~Taranenko 
  Phys.\ Rev.\  C {\bf 83}, 044902 (2011);
  R.~A.~Lacey, A.~Taranenko, N.~N.~Ajitanand and J.~M.~Alexander,
  arXiv:1105.3782 [nucl-ex].
  
\bibitem{Shen:2011zc}
  C.~Shen {\it et al.},
   J.\ Phys.\ G {\bf 38}, 124045 (2011).
  
\bibitem{Qiu:2011fi}
  Z.~Qiu and U.~Heinz,
  in {\it PANIC11}, AIP Conf. Proc., in press
  [arXiv:1108.1714 [nucl-th]].
  
\bibitem{Kestin:2008bh} 
  G.~Kestin and U.~Heinz,
  Eur.\ Phys.\ J.\ C {\bf 61}, 545 (2009).
  
\bibitem{Huovinen:2009yb}
  P.~Huovinen and P.~Petreczky,
  Nucl.\ Phys.\  {\bf A837}, 26 (2010).
  
\bibitem{Kharzeev:2001yq} 
  D.~Kharzeev, E.~Levin and M.~Nardi,
  Phys.\ Rev.\ C {\bf 71}, 054903 (2005);
  D.~Kharzeev, E.~Levin and M.~Nardi,
  Nucl.\ Phys.\ A {\bf 747}, 609 (2005).
  
\bibitem{Hirano:2005xf}
  T.~Hirano, U.~Heinz, D.~Kharzeev, R.~Lacey and Y.~Nara,
  Phys.\ Lett.\ {\bf B636}, 299 (2006).
  
\bibitem{Drescher:2006pi}
  A.~Adil, H.~J.~Drescher, A.~Dumitru, A.~Hayashigaki and Y.~Nara,
  Phys.\ Rev.\  C {\bf 74}, 044905 (2006);
  H.~J.~Drescher and Y.~Nara,
  {\it ibid.} {\bf 76}, 041903 (2007).
  
\bibitem{Hirano:2009ah}
  T.~Hirano and Y.~Nara,
  Phys.\ Rev.\  C {\bf 79}, 064904 (2009);
  and
  Nucl.\ Phys.\ {\bf A830}, 191C (2009).
  
\bibitem{Heinz:2009cv}
  U.~Heinz, J.~S.~Moreland and H.~Song,
  Phys.\ Rev.\  {\bf C80}, 061901 (2009).

\bibitem{Morita:2002av} 
  K.~Morita, S.~Muroya, C.~Nonaka and T.~Hirano,
  Phys.\ Rev.\ C {\bf 66}, 054904 (2002).

\bibitem{Vredevoogd:2012ui} 
  J.~Vredevoogd and S.~Pratt,
  arXiv:1202.1509 [nucl-th].
  
\bibitem{Bass:1998ca} 
  S.~A.~Bass
  {\it et al.},
  Prog.\ Part.\ Nucl.\ Phys.\  {\bf 41}, 255 (1998).
  
\bibitem{Song:2010aq}
  H.~Song, S.~A.~Bass and U.~Heinz,
  Phys.\ Rev.\ C {\bf 83}, 024912 (2011).

\bibitem{Qiu:2011iv}
  Z.~Qiu and U.~Heinz,
  Phys.\ Rev.\  C {\bf 84}, 024911 (2011).
  
\bibitem{Abelev:2008ez} 
  B.~I.~Abelev {\it et al.}  (STAR Collaboration),
  Phys.\ Rev.\ C {\bf 79}, 034909 (2009).

\bibitem{Back:2004dy} 
  B.~B.~Back {\it et al.}  (PHOBOS Collaboration),
  Phys.\ Rev.\ C {\bf 70}, 021902 (2004).
  
\bibitem{Alver:2008ck} 
  B.~Alver {\it et al.}  (PHOBOS Collaboration),
  Phys.\ Rev.\ C {\bf 80}, 011901 (2009).
  
\bibitem{Back:2002uc} 
  B.~B.~Back {\it et al.}  (PHOBOS Collaboration),
  Phys.\ Rev.\ C {\bf 65}, 061901 (2002).
  
\bibitem{Adler:2004zn} 
  S.~S.~Adler {\it et al.}  (PHENIX Collaboration),
  Phys.\ Rev.\ C {\bf 71}, 034908 (2005)
  [Erratum-ibid.\ C {\bf 71}, 049901 (2005)].

\bibitem{Heinz:2011kt}
  U.~Heinz, C.~Shen, and H.~Song,
  in {\it PANIC11}, AIP Conf. Proc., in press
  [arXiv:1108.5323 [nucl-th]].
  
\bibitem{Cooper:1974mv} 
  F.~Cooper and G.~Frye,
  Phys.\ Rev.\ D {\bf 10}, 186 (1974).
  
\bibitem{Shen:2011kn} 
  C.~Shen and U.~Heinz,
  Phys.\ Rev.\ C {\bf 83}, 044909 (2011).
  
\bibitem{Kolb:2000sd} 
  P.~F.~Kolb, J.~Sollfrank and U.~Heinz,
  Phys.\ Rev.\ C {\bf 62}, 054909 (2000).
  
\bibitem{Aamodt:2010pb}
  K.~Aamodt {\it et al.}  (ALICE Collaboration),
  Phys.\ Rev.\ Lett.\  {\bf 105}, 252301 (2010).

\bibitem{Aamodt:2010cz}
  K.~Aamodt {\it et al.}  (ALICE Collaboration),
  Phys.\ Rev.\ Lett.\ {\bf 106}, 032301 (2011).

\bibitem{Alt:2003ab} 
  C.~Alt {\it et al.}  (NA49 Collaboration),
  Phys.\ Rev.\ C {\bf 68}, 034903 (2003);
  S.~A.~Voloshin (STAR Collaboration),
  AIP Conf.\ Proc.\  {\bf 870}, 691 (2006).
  
\bibitem{Niemi:2011ix} 
  H.~Niemi, G.~S.~Denicol, P.~Huovinen, E.~Molnar and D.~H.~Rischke,
  Phys.\ Rev.\ Lett.\  {\bf 106}, 212302 (2011).
  
\bibitem{Heinz:2009ny} 
  U.~Heinz,
  Nucl.\ Phys.\ A {\bf 830}, 287c (2009).
  
\bibitem{Heinz:2004qz} 
  U.~Heinz,
  hep-ph/0407360.
  
\bibitem{Lisa:2000ip} 
  M.~A.~Lisa, U.~Heinz and U.~A.~Wiedemann,
  Phys.\ Lett.\ B {\bf 489}, 287 (2000).

\bibitem{Retiere:2003kf} 
  F.~Retiere and M.~A.~Lisa,
  Phys.\ Rev.\ C {\bf 70}, 044907 (2004).

\bibitem{Barrette:1997fj} 
  J.~Barrette {\it et al.}  (E877 Collaboration),
  Phys.\ Rev.\ Lett.\  {\bf 78}, 2916 (1997).

\bibitem{Lisa:2000hw} 
  M.~A.~Lisa {\it et al.}  (E895 Collaboration),
  Phys.\ Rev.\ Lett.\  {\bf 84}, 2798 (2000).

\bibitem{Adams:2004yc} 
  J.~Adams {\it et al.}  [STAR Collaboration],
  Phys.\ Rev.\ C {\bf 71}, 044906 (2005).
    
\bibitem{Abelev:2009tp} 
  B.~I.~Abelev {\it et al.}  (STAR Collaboration),
  Phys.\ Rev.\ C {\bf 80}, 024905 (2009).
  
\bibitem{Lisa:2011na} 
  M.~A.~Lisa, E.~Frodermann, G.~Graef, M.~Mitrovski, E.~Mount, H.~Petersen and M.~Bleicher,
  New J.\ Phys.\  {\bf 13}, 065006 (2011).

\bibitem{Anson:2011ik} 
  C.~Anson  {\it et al.} (STAR Collaboration),
  J.\ Phys.\ G {\bf 38}, 124148 (2011).
  
\bibitem{Heinz:2002sq} 
  U.~Heinz and P.~F.~Kolb,
  Phys.\ Lett.\ B {\bf 542}, 216 (2002).

\bibitem{Frodermann:2007ab} 
  E.~Frodermann, R.~Chatterjee and U.~Heinz,
  J.\ Phys.\ G G {\bf 34}, 2249 (2007).

\end{thebibliography}

\end{document}